\documentclass[journal,twocolumn]{IEEEtran}
\usepackage{amsfonts}
\usepackage{times}
\usepackage{graphicx}
\usepackage{latexsym}
\usepackage{dsfont}
\usepackage{amssymb}
\usepackage{amsmath}
\usepackage{cite}
\usepackage{verbatim}
\usepackage{subfigure}

\def\bb0{{\mathbb{0}}}

\def\bb{{\mathbf{b}}}

\def\bp{{\mathbf{p}}}

\def\b0{{\mathbf{0}}}

\def\bS{{\mathbf{S}}}

\def\bX{{\mathbf{X}}}

\def\sf0{{\mathsf{0}}}

\usepackage{multirow}
\usepackage{multicol}
\usepackage{booktabs}
\usepackage{epstopdf}
\usepackage{enumerate}
\usepackage{algorithm}
\usepackage{amsmath}
\usepackage{float}
\usepackage{color}
\usepackage{makeidx}
\usepackage{bm}
\usepackage{cleveref}
\usepackage{url}
\usepackage{steinmetz}
\usepackage{varwidth}% http://ctan.org/pkg/varwidth
\usepackage{soul}
\graphicspath{{figs/}}
\usepackage{tabularx}
\usepackage{comment}
\usepackage{lipsum}
\usepackage{subfigure}
\usepackage{balance}
\newcommand{\comm}[1]{}
\begin{document}

	\title{Environment Semantic Communication: \\Enabling Distributed Sensing Aided Networks \\ }
	\author{Shoaib Imran, Gouranga Charan, and Ahmed Alkhateeb\\ \thanks{The authors are with the School of Electrical, Computer, and Energy Engineering, Arizona State University. Emails: \{s.imran, gcharan, alkhateeb\}@asu.edu. This work is supported in part by the National Science Foundation under Grant No. 2048021. Part of this work has been accepted in the IEEE International Conference on Communications Workshops \cite{10283602}. }}
	
	\maketitle

	\begin{abstract}
		Millimeter-wave (mmWave) and terahertz (THz) communication systems require large antenna arrays and use narrow directive beams to ensure sufficient receive signal power. However, selecting the optimal beams for these large antenna arrays incurs a significant beam training overhead, making it challenging to support applications involving high mobility. In recent years, machine learning (ML) solutions have shown promising results in reducing the beam training overhead by utilizing various sensing modalities such as GPS position and RGB images. However, the existing approaches are mainly limited to scenarios with only a single object of interest present in the wireless environment and focus only on co-located sensing, where all the sensors are installed at the communication terminal. This brings  key challenges such as the limited sensing coverage compared to the coverage of the communication system and the difficulty in handling non-line-of-sight scenarios. To overcome these limitations, our paper proposes the deployment of multiple distributed sensing nodes, each equipped with an RGB camera. These nodes focus on extracting environmental semantics from the captured RGB images. The semantic data, rather than the raw images, are then transmitted to the basestation. This strategy significantly alleviates the overhead associated with the  data storage and  transmission of the raw images. Furthermore, semantic communication enhances the system's adaptability and responsiveness to dynamic environments, allowing for prioritization and transmission of contextually relevant information. Experimental results on the DeepSense 6G dataset demonstrate the effectiveness of the proposed solution in reducing the sensing data transmission overhead while accurately predicting the optimal beams in realistic communication environments. 
	\end{abstract}
	
	\begin{IEEEkeywords}
		Millimeter-wave, environment semantics, semantic communications, distributed sensing, camera, deep learning, computer vision, beamforming. 
	\end{IEEEkeywords}

	\section{Introduction}\label{sec:intro}
	
	Utilizing higher frequency bands, such as mmWave in 5G and possibly sub-terahertz in 6G, is a key trend in current and future communication systems. 
	These frequency ranges provide higher bandwidths, enabling the communication systems to efficiently meet the higher data rate demands of emerging applications such as augmented/virtual reality, autonomous vehicles, and smart cities \cite{alkhateeb2018deep,rappaport2019wireless, 8300313}. 
	However, these systems necessitate the deployment of large antenna arrays and the use of narrow beams at both the transmitter and receiver to ensure adequate receive signal power. 
	Selecting the best beams for these large antenna arrays incurs a substantial training overhead, making it challenging to satisfy the low-latency and high-reliability requirements of these current and future applications. This emphasizes the need to explore innovative approaches that (i) reduce or mitigate the training overhead associated with beam selection and (ii) enable highly mobile wireless communication applications.
	
	Several solutions have been proposed over the years to reduce the beam training and channel estimation overhead in mmWave communication systems \cite{6162448, Alkhateeb2014, 8025577, 9289508}. The focus of these solutions has been mainly on: (i) The development of beam training with adaptive/hierarchical beam codebooks \cite{Hur2011, Alkhateeb2014, 6162448}. (ii) The utilization  of compressive sensing tools \cite{Alkhateeb2014, 6847111} to estimate the full channel with a much smaller number of measurements. This is motivated by the sparsity nature of the mmWave channels, where only a few dominant paths typically exist between  the transmitter and receiver.  (iii) The design of beam tracking techniques \cite{8025577} that leverages the user mobility information to predict the future beams and hence reduce the exhaustive search beam training overhead. These classical approaches, however, usually result in a training overhead reduction of only one order of magnitude, which is not sufficient for very large antenna array systems and applications that require very low-latency.

	\begin{figure*}
		\centering
		\includegraphics[width=1\linewidth]{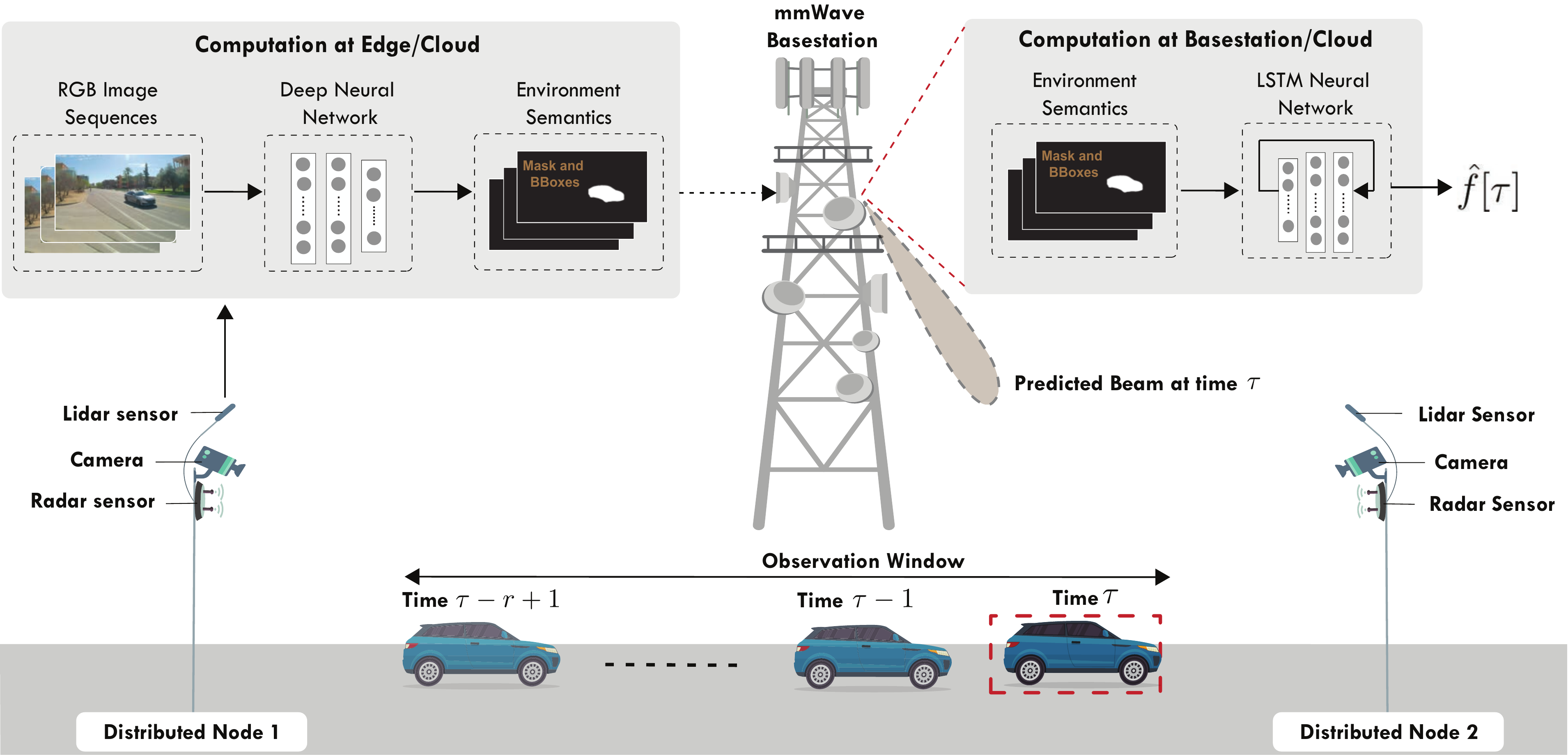}
		\caption{The figure shows the overall system model of the proposed setup. The distributed nodes extract environment semantic information from the RGB images, which is subsequently transmitted to the basestation. This semantic information is then utilized for beam prediction at the basestation.}
		\label{fig:sys_model1}
	\end{figure*}
	
	The challenges faced by classical solutions have led to the development of machine learning approaches that leverage prior observation and additional sensing information \cite{alkhateeb2023real, Morais22, charan2021c, charan2023camera, jiang2023, demirhan2021beam, demirhan2023integrated}. The additional sensing modalities include position (GPS location) \cite{Morais22}, RGB images \cite{charan2021c, charan2023camera}, LiDAR \cite{jiang2023}, and Radar \cite{demirhan2021beam, demirhan2023integrated}. The additional sensing information provides a crucial environmental context, enabling an in-depth comprehension of the wireless environment and its influence on channel characteristics. These prior studies have demonstrated the potential of utilizing additional side information in minimizing the beam training overhead. However, these solutions have certain limitations. Firstly, they are primarily designed for scenarios with a single object of interest, which can be challenging when scaling them to real-world situations with multiple objects. Secondly, the additional sensors used in these solutions, such as cameras, LiDAR, and radar, are positioned exclusively at the basestation and have a limited range of approximately $60-80$ meters \cite{lotscher2023assessing}. This range is significantly shorter than the typical range of the mmWave communication systems, which is around $300$ meters. Consequently, this limited range of these additional sensing modalities significantly impacts the effectiveness of these solutions in real-world wireless communication tasks (such as beam prediction and proactive blockage prediction). Additionally, these sensors do not provide coverage for non-line-of-sight scenarios, further restricting their applicability in diverse environments.
	
	One promising solution to overcome these challenges is deploying multiple nodes, each equipped with its own sensors, to capture information about the wireless environment in a coordinated manner. This distributed sensing approach enhances coverage, reliability, and adaptability by strategically distributing sensors throughout the network \cite{park2021communication, xu2023edge}. Instead of relying solely on sensors at the basestation, data collected by these distributed nodes can be utilized by one or more basestations to make informed decisions. This scalable and robust approach leverages the collective sensing capabilities of multiple nodes, providing a comprehensive view of the wireless environment and optimizing tasks such as beam prediction and proactive blockage prediction. However, as the number of distributed nodes increases, challenges arise in managing the growing volume of captured data, including storage, processing, and transmission concerns. Furthermore, the heightened data rate resulting from the increased number of nodes necessitates robust data synchronization methods to maintain temporal coherence. One way to address these challenges is by processing the data captured by the distributed nodes locally, either at the edge or in the cloud. This involves extracting essential information, referred to as ``environment semantics", from the raw sensor data. Environment semantics encompasses meaningful details about the wireless environment, which includes the number, type, and shape of the objects, among other relevant attributes. As such, these environment semantics have the potential to accurately represent the information within the wireless environment while also minimizing the data storage requirement as compared to the original sensing modality.
	
	While exploring distributed learning solutions, it is notable that paradigms such as federated learning \cite{niknam2020federated, yang2020energy} are gaining traction. This method entails training models across decentralized devices using local data, with the aggregated insights refining the overall model while maintaining data privacy and minimizing bandwidth use. In addition, distributed artificial intelligence (AI) and edge computing have shown potential in enhancing network functionalities. These technologies, particularly edge computing, have been effective in managing the data from distributed nodes, addressing significant challenges in data processing and storage \cite{wang2019edge, li2019edge}. However, these advancements still need to be explored in the context of distributed sensing-aided vehicle-to-infrastructure (V2I) beam prediction and tracking. To bridge this gap, we aim to investigate the utilization of environment semantics in enabling distributed sensing-aided wireless communication in real-world scenarios. Specifically, we propose a novel approach that leverages environment semantics in a distributed sensing scenario to predict optimal beams in a real-world wireless communication setting accurately. The main contributions of this paper can be summarized as follows:
	\begin{itemize}
		\item Formulating the sensing-aided beam prediction problem for vehicle-to-infrastructure (V2I) communication scenario with multiple distributed nodes, each equipped with an RGB camera to capture the wireless environment. 
		
		\item Developing a novel deep learning-based solution that leverages images captured by cameras installed at distributed nodes to accurately predict the optimal beam index at the basestation in a V2I communication scenario.
		
		\item Investigating various environment semantics that can be extracted from images, such as object bounding boxes and masks, with the aim of enabling distributed sensing-aided wireless communication. We further perform a comprehensive comparative study, evaluating the performance, complexity, and practical feasibility of these environment semantics for the specific task of distributed sensing-aided beam prediction.
		
		\item Providing the first real-world evaluation of distributed environment semantic-aided beam prediction based on a new scenario in the DeepSense 6G dataset \cite{DeepSense}. This scenario focuses explicitly on the distributed aspect, capturing co-existing multi-modal data from the basestation and two distributed units, offering a comprehensive dataset that enables the study of distributed sensing-aided wireless communication.
		
	\end{itemize}

	The paper is organized as follows: Section \ref{sec:sysmodel_and_probform} provides the system model and problem formulation for the proposed solution. In Section \ref{sec:keyidea} and Section \ref{sec:propsol}, we delve into the key idea and the proposed solution, respectively. The testbed and the DeepSense 6G dataset utilized in our experiments are described in Section \ref{sec:testbed}. Finally, in Section \ref{sec:perf_eval} we present a detailed evaluation of the proposed solution.
	
	%####################################################################################
	%####################################################################################
	%####################################################################################

	\section{System Model and Problem Formulation}\label{sec:sysmodel_and_probform}
	In this section, we present the adopted system model and formulate the distributed environment semantic-aided beam prediction problem.
	
	\subsection{System Model}\label{sec:Sysmodel}
	Fig. \ref{fig:sys_model1} illustrates the proposed distributed sensing-aided communication setup. In this setup, $N$ distributed nodes sense the environment and transmit environment semantic information to a basestation that is serving a mobile user. Each distributed node in the system is assumed to be equipped with an RGB camera. Furthermore, the base station is equipped with an RGB camera and 3 $M$-element uniform linear arrays (ULAs), with each ULA having a field of view of around $90^\circ$. The three uniform linear arrays (ULAs) are positioned $90^\circ$ apart from each other and oriented towards the front, left, and right of the basestation. Furthermore, the area served by the basestation is divided into $N+1$ subregions. The basestation camera provides sensing information for the region directly in front of the basestation while each distributed node provides sensing information for one of the remaining $N$ regions. We strategically position the distributed nodes to enhance the combined camera coverage over the range of the mmWave communication system. 
	
	The user is equipped with a single-antenna transmitter and a GPS receiver for collecting real-time position information. The basestation, for each ULA, uses (i) OFDM transmission with $K$ subcarriers and a cyclic prefix of length $D$, and (ii) a pre-defined beam steering codebook $\boldsymbol{\mathcal F}=\{\mathbf f_q\}_{q=1}^{Q}$, where  $\mathbf{f}_q \in \mathbb C^{M\times 1}$ is the $q^{th}$ beamforming vector and $Q$ is the total number of beamforming vectors.  The beam steering beams are uniformly spaced and jointly cover the ULA's $90^\circ$ field of view. In the downlink, the received signal at the user from the ULA that has the user in its field of view at the $k^{th}$ sub-carrier and time $t$ can be represented as
	\begin{equation}\label{eq:t1}
		y_{k}[t] = \mathbf h_{k}^T[t] \mathbf f_q[t]x + v_k[t],
	\end{equation}
	where  $\mathbf h_{k}[t] \in \mathbb C^{M\times 1}$ denotes the channel between the basestation and the mobile user,  $\mathbf f_q \in \boldsymbol{\mathcal F}$ is the beamforming vector,  and $v_k[t]$ represents noise sampled from a complex Gaussian distribution $\mathcal N_\mathbb C(0,\sigma^2)$. The transmitted complex symbol $x\in \mathbb C$  satisfies the power constraint $\mathbb E\left[ |x|^2 \right] = P$, where $P$ is the average symbol power. Moreover, the beamforming vector  $\mathbf f_q[t]$, at each time step $t$ is selected from the beam steering codebook $\boldsymbol{\mathcal F}$ to maximize the average receive SNR as follows
	\begin{equation}\label{t2}
		\underset{\mathbf f_q[t]\in \boldsymbol{\mathcal F}}{\text{argmax}} \frac{1}{K}\sum_{k=1}^{K} \mathsf{SNR}|\mathbf h_{k}^T[t] \mathbf f_q[t] |^2,
	\end{equation}
	where $\mathsf{SNR}$ is the transmit signal-to-noise ratio, $\mathsf{SNR} = \frac{P}{\sigma^2}$.  At any time instant $t$, the receive power vector of effective channel gain with codebook elements from the ULA that has the user in its field of view can therefore be expressed as $\mathbf{p}[t] = [p_1[t],...,p_Q[t]]$, where $\mathbf{p}[t] \in \mathbb R^{Q\times 1}$ and $p_q[t]$ is defined as 
	\begin{equation}\label{t3}
		p_q[t] = |\mathbf{h}_k^T[t]\mathbf{f}_q[t]|^2. \quad q \in {1, ....., Q}
	\end{equation}
	In the next subsection, we formulate the distributed environment semantic-aided beam prediction problem.
	
	\subsection{Problem Formulation}\label{Sec:prob_form}
	Given the system model presented above, the goal is to select the optimal beam index (at the basestation for any given time $t$) that maximizes the receive power using camera images captured by the distributed node. Attaining this goal involves a few key tasks. Firstly, we need to determine the ULA that encompasses the user within its field of view. Secondly, we have to identify the sub-region where the user is located. Lastly, we must also discern the transmitter vehicle from other vehicles present in the RGB images. One potential solution to this problem is to leverage the receive power vector derived from previous time instances. Consequently, this work aims to develop a beam prediction model that utilizes a sequence of available RGB images and the ground truth receive power vector corresponding to the time instant of the first image capture. The receive power vector corresponding to the first image capture in the sequence serves a dual role in our proposed solution. Initially, it is used to identify the ULA and the sub-region where the user is located. Subsequently, this receive power vector is employed to facilitate the identification of the transmitter in the scene. Let $\mathbf{X}_n[t] \in \mathbb{R}^{W \times H \times C}$ represent the RGB image captured at time $t$ by the camera installed at the $n^{th}$ node, where $W$, $H$, and $C$ are the width, height, and the number of color channels for the image, respectively. Further, let  $\bp[t] \in \mathbb{R}^{1 \times Q}$ denote the mmWave receive power vector from the ULA that has the user in its field of view at time $t$. At any given time instant $t$, the distributed node $n$, captures a sequence of $r$ RGB images, and the basestation collects the mmWave receive power vector corresponding to the time instant of the first image capture, $\bS[t]$, defined as
	\begin{equation}
		{\bS}[t] = \left\{ \left\{ \bX_n[t] \right\}_{t = \tau-r+1}^{t= \tau}, \bp[\tau-r+1] \right\}, 
	\end{equation}
	where $r \in \mathbb Z$ is the length of the input sequence or the observation window to predict the optimal beam index. In particular, at any given time instant $t$, the goal in this work is to find a mapping function $f_{\Theta}$ that utilizes the available sensory data samples $\bS[t]$ to predict (estimate) the optimal beam index $ \hat{\mathbf f}[t] \in \boldsymbol{\mathcal F}$ with high fidelity. The mapping function can be formally expressed as
	\begin{equation}
		f_{\Theta}: \bS[t] \rightarrow  \hat{\mathbf f}[t].
	\end{equation}
	
	Let $\mathcal{D} = \left\lbrace (\bS_{l}, \mathbf{f}^\star_{l}) \right\rbrace_{l=1}^{l=\varkappa_1}$ represent the available dataset collected from the real-world wireless environment. The total number of samples in the dataset is denoted by $\varkappa_1$. The goal is to maximize the number of correct predictions over all the samples in the dataset $\mathcal{D}$. This can be formally written as:
	\begin{equation}\label{u1b}
		f^{\star}_{\theta^{\star}} = \underset{f_{\theta}}{\text{argmax}} \prod_{l=1}^{\varkappa_1} \mathbb P\left( \hat{\mathbf{f}}_l = \mathbf f_l^{\star} | \bS_{l} \right),
	\end{equation}
	where the joint probability distribution in \eqref{u1b} is due to the implicit assumption that the samples in $\mathcal{D}$ are drawn from an independent and identical distribution. The objective is to find the optimal set of parameters $\Theta^\star$ that maximizes the product of the probabilities of correct predictions. The next section presents the proposed deep learning-based solution for the proposed distributed sensing-aided beam prediction.
	
	%%%%%%%%%%%%%%%%%%%%%%%%%%%%%%%%%%%%%%%%%%%%%%%%%%%%%%%%%%%%%%%%%%%%%%%%%%%%%%%%%%%%%%%%%%%%%%%%%%%%%%%%%%%%%%%%%%%%
	
	\begin{figure}
		\centering
		\includegraphics[width=1\linewidth]{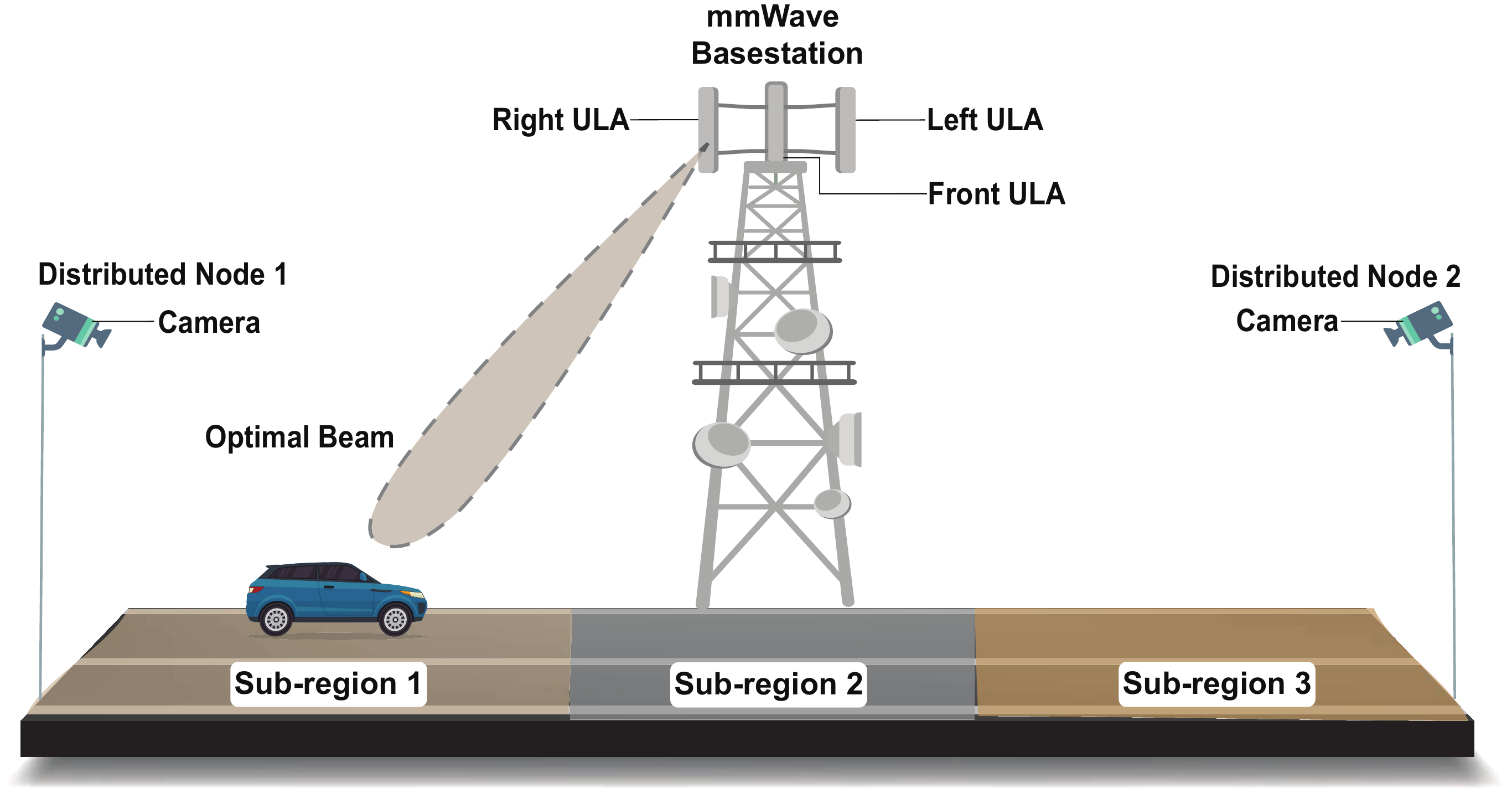}
		\caption{The figure illustrates the selection process of the ULA, the sub-region, and the corresponding distributed node.}
		\label{fig:ULA_selection}
	\end{figure}

	\section{Key Idea}\label{sec:keyidea}
	In this section, we present the key idea behind setting up distributed nodes and utilizing the environment semantics from these distributed nodes for beam prediction at the basestation. 
	
	Recent works \cite{Morais22, charan2021c, charan2023camera, jiang2023, demirhan2021beam, demirhan2023integrated} demonstrate the potential of using various sensing modalities including position (GPS location) \cite{Morais22}, LiDAR \cite{jiang2023}, radar \cite{demirhan2021beam}, and RGB images \cite{charan2021c} for beam prediction. These works primarily focus on co-located sensing and communication, where sensors are installed at the basestation. This approach introduces various challenges that need to be addressed. First, practical sensors have limited range capabilities. This discrepancy in range poses a challenge for ensuring seamless integration and synchronization of data between communication and sensing modalities. Furthermore, when these models do detect distant objects, the resulting bounding boxes and masks may appear disproportionately small, giving the impression of minimal movement within the image, even if the objects are actually moving rapidly. Second, the sensors predominantly rely on line-of-sight (LOS) conditions for accurate data capture \cite{wu2023proactively}. Consequently, non-line-of-sight (NLOS) scenarios, which are inherently challenging even for mmWave communication systems, significantly impact the usability of these sensors. Achieving reliable mmWave communication necessitates the development of techniques that can effectively handle both LOS and NLOS cases, ensuring robust and uninterrupted communication in various real-world scenarios.

	\begin{figure*}
		\centering
		\includegraphics[ width=1\linewidth]{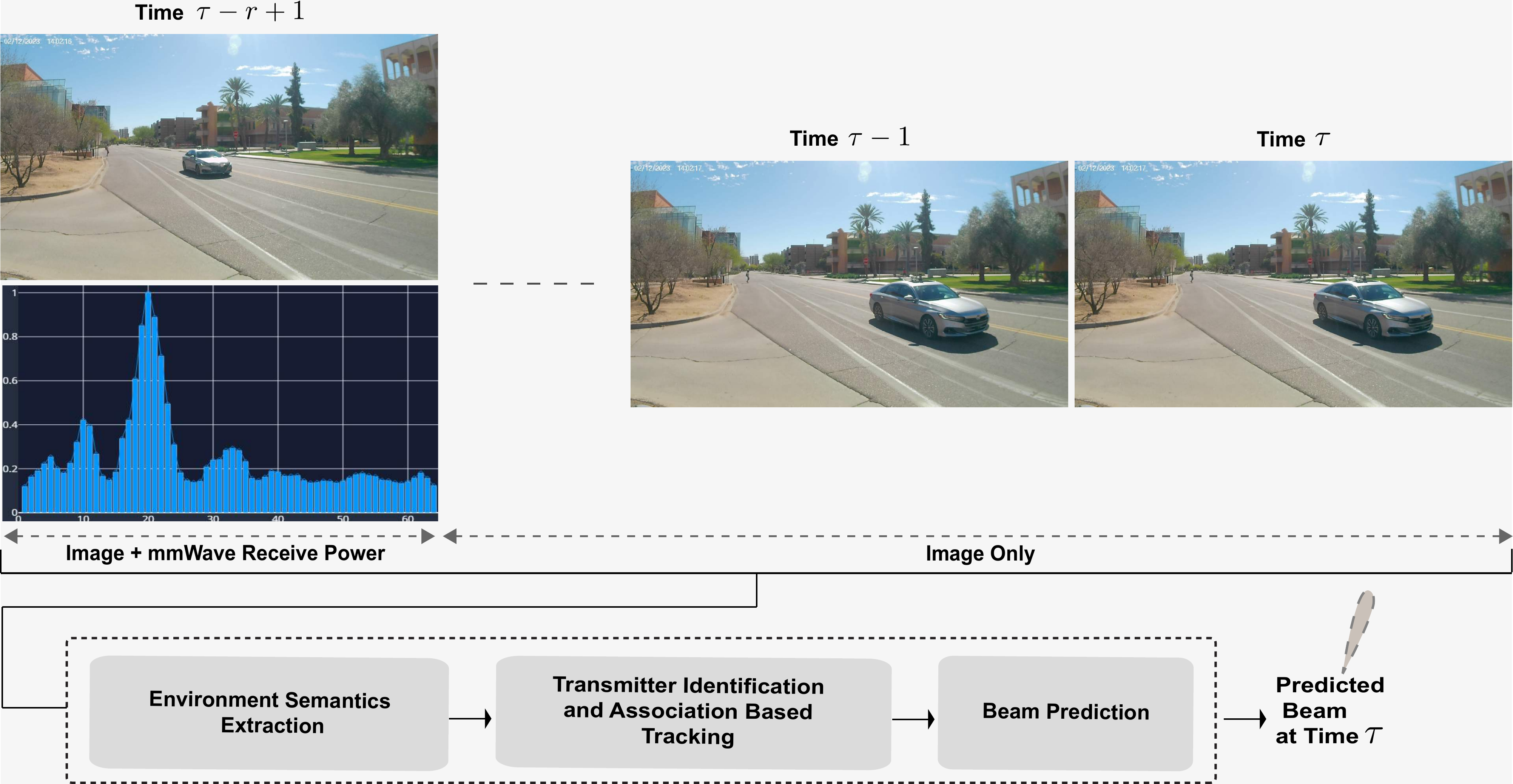}
		\caption{This figure outlines the different stages of the proposed solution. First, we extract environment semantics from the raw RGB images, transmitting them to the basestation. In the second stage, we identify the transmitter in the initial frame and track it over the subsequent frames. The final step involves using this semantic information of the transmitter, accumulated in the second stage, for beam prediction.}
		\label{fig:prop_sol}
	\end{figure*}
	
	To unlock the full potential of additional sensing modalities, \textbf{adopting a distributed sensing approach is essential}. This approach entails deploying multiple distributed nodes, each equipped with sensors such as camera, LiDAR, and radar. By distributing the sensors, we can overcome limitations in range and expand the scope of data collection to cover a broader area. Furthermore, distributed sensing enables us to address NLOS scenarios effectively by enhancing sensing coverage and capturing diverse perspectives. By capitalizing on the synergistic capabilities of multiple sensors deployed across the network, we can elevate overall system performance and realize advanced functionalities.
	
	As the number of distributed nodes increases, there is a \textbf{corresponding increase in the volume of captured data}, which brings forth several challenges that necessitate careful consideration. Firstly, the substantial size of the accumulated data presents significant obstacles in terms of storage, processing, and transmission. Effectively managing and efficiently storing the larger data volume requires implementing robust storage solutions and high-speed data processing capabilities to ensure good system performance. Secondly, the heightened data rate resulting from the increased number of nodes calls for robust data synchronization mechanisms to maintain temporal coherence and mitigate potential data inconsistencies. Synchronizing the data collected at the distributed nodes with the basestation's central processing unit is essential to ensure accurate analysis. Addressing these challenges associated with the growing scale of data capture and synchronization overhead is paramount to facilitate seamless operation and enable the effective utilization of distributed sensing techniques in mmWave communication systems.
	
	One promising approach to overcome these challenges is reducing the data traffic volume between the basestation and the distributed nodes by selectively \textbf{transferring only critical information}. For instance, in the case of a distributed node equipped with a camera, rather than transmitting the entire image, a more efficient strategy is to extract the environment semantics locally. These environment semantics comprise relevant information about the wireless environment, such as the presence of different vehicles in the scene and their relative locations. By focusing on transmitting only this critical information, the data traffic between the distributed nodes and the basestation can be significantly reduced, alleviating the storage, processing, and transmission burdens. This approach allows for the efficient utilization of network resources while maintaining the necessary level of information for tasks such as V2I beam prediction. By extracting and transmitting only the essential details, the overall system performance can be enhanced, effectively addressing the challenges associated with the growing volume of data traffic in distributed sensing-based mmWave communication systems.
	
	To address the challenges mentioned above, this paper focuses on designing efficient strategies for extracting \textbf{environment semantics from RGB images (such as the masks and bounding boxes of the objects of interest)} to facilitate the beam prediction process in a realistic V2I communication scenario with two distributed nodes. In this scenario, we consider a more complex environment with multiple probable objects, requiring advanced techniques for accurate prediction. Given the multi-candidate nature of the scenario, our solution is designed to efficiently extract environment semantics, identify the transmitter in the scene, and predict the optimal beam in real time. To accomplish this, we leverage a sequence of RGB images captured by the distributed nodes. However, one key challenge that still remains is the latency associated with transferring these environment semantics to the basestation for beam prediction. It is important to note here that our solution can also be extended to predict future beams, enhancing the proactive nature of the system. By adopting a proactive approach and predicting future beams, we can effectively overcome the latency issue, ensuring timely and accurate beam prediction in distributed sensing-based mmWave communication systems. Next, we present the proposed solution in detail. 
	
	%%%%%%%%%%%%%%%%%%%%%%%%%%%%%%%%%%%%%%%%%%%%%%%%%%%%%%%%%%%%%%%%%%%%%%%%%%%%%%%%%%%%%%%%%%%%%%%%%%%%%%%%%%%%%%%%%%%%%%%%%%%%%%%%%%%%%%%%%%

	\begin{figure*}
		\centering
		\includegraphics[width=1\linewidth]{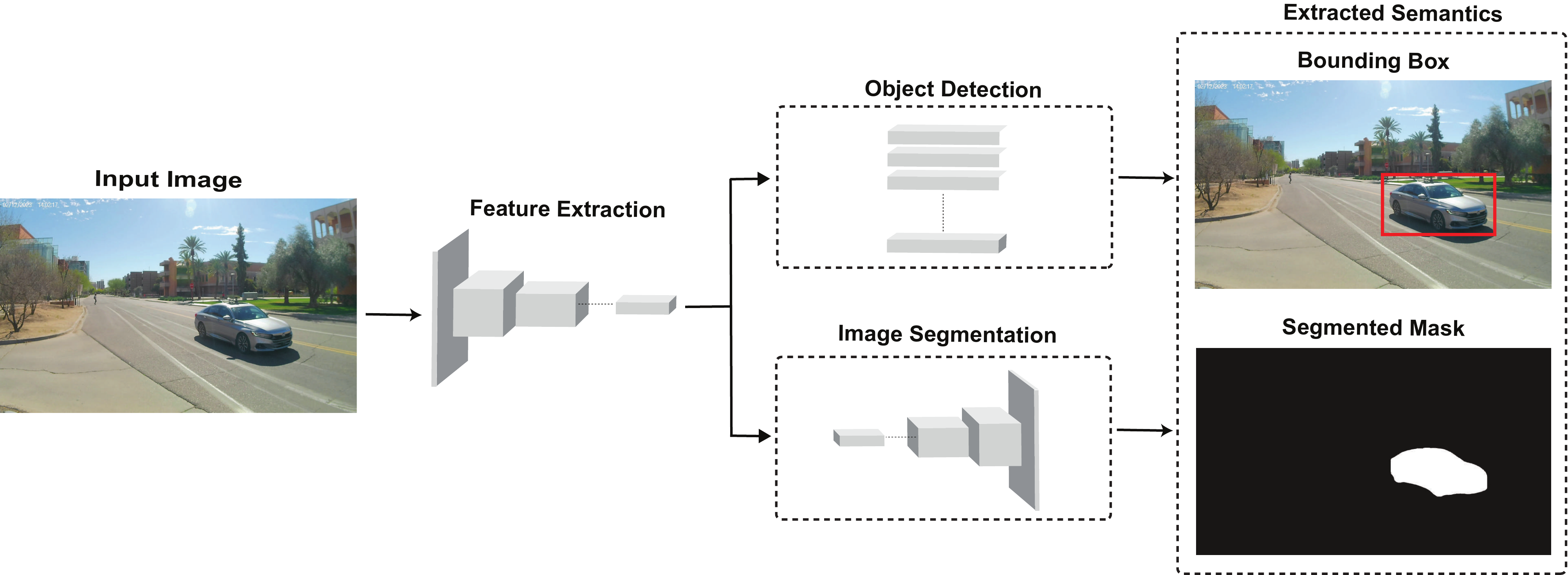}
		\caption{The figure illustrates the environment semantics extraction stage in our proposed solution. In particular, a camera installed at the distributed node captures real-time images of the wireless environment, which a machine learning model then processes to extract the bounding boxes and masks of the mobile objects present in the images.}
		\label{fig:env_sem_extract}
	\end{figure*}
	
	\section{Proposed Solution}\label{sec:propsol}
	This section provides a comprehensive overview of the proposed solution for distributed environment semantic-aided communication. In order to simplify the analysis, we divide the region served by the basestation into three sub-regions, where each sub-region corresponds to one of the phased arrays of the basestation as shown in Fig. \ref{fig:ULA_selection}. Furthermore, we include two distributed nodes in the system with one node located to the left of the basestation and the other to the right. At any given time $t$, the selection of sensing data for further processing and beam prediction depends on the user's location in the wireless environment. For instance, if the user is situated in the sub-region to the right of the basestation, the RGB images captured by the right distributed node (distributed node 1) are utilized for beam prediction. It is important to note here that the selection of the distributed node for further processing and beam prediction does not explicitly rely on the user's position data (GPS position). Instead, we utilize the receive power vector, which provides valuable directional information that aids in determining the optimal beam index. By utilizing the optimal beam index, we can select one of the ULAs. Next, depending on the selected ULA, we approximate the sub-region where the user is located. This approximation, in turn, helps identify the specific distributed node from which to utilize the sensing data. This approach ensures efficient utilization of the sensing data (available from the distributed nodes) based on the user's location within the coverage area.

	The subsequent solution comprises three stages as depicted in Fig. \ref{fig:prop_sol}. In the first stage, an environment semantics extraction process is employed, utilizing a machine learning model to extract object masks and bounding boxes of potential users. It is worth noting that our focus in this work is primarily on mobile vehicles within the context of vehicle-to-infrastructure communication. In the second stage, we utilize the receive power vector to identify and track the transmitter over the subsequent $r-1$ frames using the nearest neighbor algorithm. Moreover, we explore how incorporating semantic information, such as the color of vehicles, can enhance the accuracy of object association-based tracking. Lastly, in the third stage, the basestation leverages the transmitter's semantic information from the current and past $r$ frames to predict the current optimal beam index. By incorporating semantic information about the transmitter and utilizing receiver power data, our proposed approach aims to predict the optimal beam index accurately.

	\subsection{Stage 1: Environment Semantics Extraction}\label{subsec:env_sem_extract}
	The first stage of the proposed solution aims to extract environment semantics from RGB images, as shown in Fig. \ref{fig:env_sem_extract}. The primary objective of this stage is to accurately and efficiently capture information that represents the objects of interest in the wireless environment while also minimizing the required data storage compared to the original sensing modality (i.e., the images themselves). To achieve this, we utilize the state-of-the-art  COCO \cite{lin2014microsoft} pre-trained object detection and image segmentation model, YOLOv7 \cite{wang2022yolov7}. By leveraging YOLOv7, we aim to extract two crucial types of environment semantics: bounding boxes and binary masks. Bounding boxes, denoted as $\mathbf{X}_{\text{BBox}}[t] \in \mathbb{R}^{U \times 4}$, serve as representations for potential users within the wireless environment, where $U$ is the total number of detected objects in the RGB image. Each row of $\mathbf{X}_{\text{BBox}}[t]$ contains a bounding box vector $[x_c, y_c, w, h]$, where $x_c$, $y_c$, $w$, and $h$ denote the $x$-center, $y$-center, width, and height of the detected object, respectively. These bounding boxes provide essential spatial information about the potential users. Additionally, we generate binary masks, represented as $\mathbf{X}_{\text{Mask}}[t] \in \mathbb{R}^{\hat{W} \times \hat{H}}$, where $\hat{W}$ and $\hat{H}$ correspond to the downsampled width and height of the image mask, respectively. These masks offer more detailed and fine-grained depictions of the spatial extent of the potential users within the wireless environment. Furthermore, the image segmentation model employed in our solution not only outputs the binary masks but also provides the bounding box information for the detected objects. Let $\mathbf{X}_{\text{B-Mask}}[t] \in \mathbb{R}^{U \times 4}$ denote the bounding boxes extracted during the image segmentation.

	\subsection{Stage 2: Transmitter Identification and Tracking} \label{subsec:txid_track}
	By adopting YOLOv7, which allows the simultaneous generation of bounding boxes and masks, we eliminate the need for separate runs, resulting in faster inference speed. Moreover, YOLOv7 achieves a significant reduction of approximately $40\%$ in parameter size compared to other real-time object detectors, leading to enhanced computational efficiency. Utilizing pre-trained models based on the COCO dataset is advantageous as they can detect most of the relevant objects commonly encountered in wireless environments, such as cars, bikes, and people. By extracting environment semantics, we strive to accurately represent the objects of interest while reducing the data storage requirements compared to the original RGB images. This approach enables more efficient processing and transmission of critical information between the distributed nodes and the basestation, ultimately improving overall system performance and alleviating computational burdens. The next stage of the proposed solution is to identify the transmitter in the RGB image. 
	
	\begin{figure*}
		\centering
		\includegraphics[width=1\linewidth]{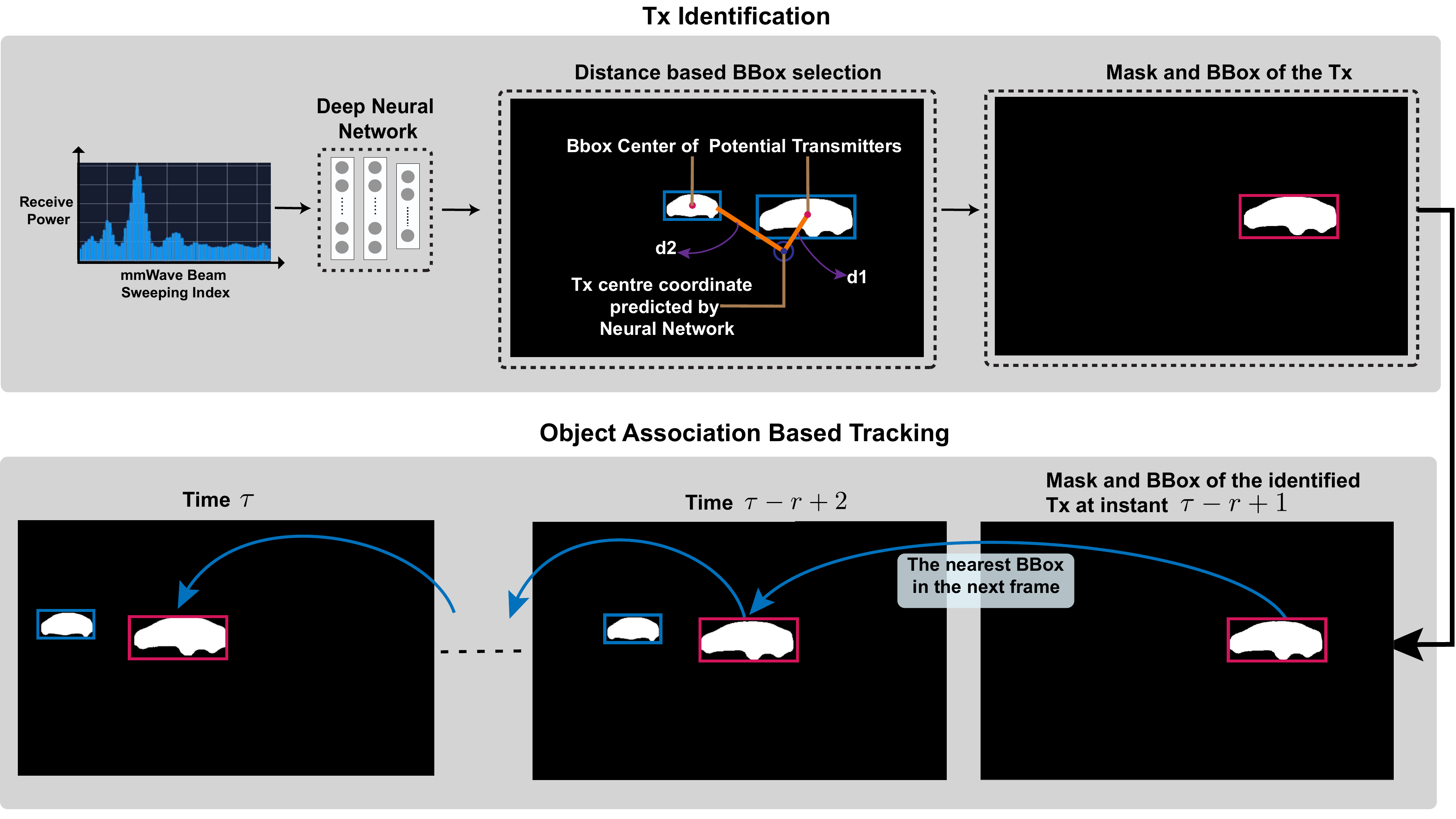}
		\caption{The figure shows the transmitter identification and object association-based tracking module. The transmitter is identified in the first frame using the receive power vector and then tracked for the remaining frames using the nearest neighbor algorithm.}
		\label{fig:trans_iden}
	\end{figure*}
	
	Building upon the environment semantics extracted in the previous stage, which provides information about all relevant objects in the scene, we now present the second stage of our proposed solution. This stage plays a crucial role in our objective of predicting the optimal beam index in a real-world wireless environment with multiple potential users. Specifically, two tasks need to be accomplished to predict the optimal beam index accurately. Firstly, the transmitter must be identified among the detected objects, and secondly, the transmitter needs to be tracked over the subsequent $r-1$ samples. To address these challenges, we introduce a two-stage solution encompassing transmitter identification and object association-based tracking, as presented in Fig. \ref{fig:trans_iden}. These stages work together to accurately predict the optimal beam index. In the following sections, we provide a comprehensive description of our proposed transmitter identification and object association-based tracking solution.

	\subsubsection{Transmitter Identification} It refers to accurately determining the transmitter's location within the wireless environment using the extracted semantic information, such as bounding boxes and masks. This subsection presents the second stage of our proposed solution, which focuses on transmitter identification. As mentioned earlier, the adopted image segmentation model not only provides binary masks but also supplies the bounding box information for the detected objects. In the task of transmitter identification, we leverage the extracted bounding boxes from both types of environment semantics. Therefore, the objective is to leverage the  receive power vector $\mathbf{p}[\tau-r+1]$ from the ULA that has the user in its field of view and the semantic information of masks and bounding boxes at time $t=\tau-r+1$ to predict the center coordinates of the transmitter's bounding box $\mathbf{b}_\text{Tx}[\tau -r +1] \in \mathbb{R}^{2 \times 1}$ within the image. For this, we employ a prediction function $g_\eta$, parameterized by a set of parameters $\eta$, which maps the receive power vector to the predicted bounding box center coordinates $\mathbf{\hat{b}}_{\text{Tx}}$. Mathematically, this can be expressed as: 
	\begin{equation}
		\label{t6} 
		g_\eta: \mathbf{p}[t] \rightarrow \hat{\mathbf{b}_\text{Tx}}[t]. 
	\end{equation}
	To train the prediction function, we construct a dataset $\mathcal{D}_2$ comprising pairs of mmWave receive power vectors ${\mathbf{p}}_v$ and their corresponding ground-truth bounding box center coordinates of the transmitter $ {\mathbf{b}_{\text{Tx}}}_v$. This dataset is a subset of the larger dataset $\mathcal{D}$, and it contains $V$ samples, such that $\mathcal{D}_2 = \left \lbrace ({\mathbf{p}}_v, {\mathbf{b}_{\text{Tx}}}_v) \right \rbrace _{v=1}^{V}$. The goal is to minimize the error between the predicted and ground-truth center coordinates of the transmitter's bounding box across all the samples in $\mathcal{D}_2$. This optimization problem can be formulated as: 
	\begin{equation} 
		g^{\star}_{\eta^\star} = \underset{g_\eta}{\text{argmin}} \frac{1}{V}\sum_{v=1}^{V} \| {\hat{\mathbf{b}_{\text{Tx}}}}_v - {\mathbf{b}_{\text{Tx}}}_v\|^2, 
	\end{equation} 
	where $g^{\star}_{\eta^\star}$ represents the optimal prediction function that minimizes the squared $l_2$ norm of the error between the predicted and ground-truth bounding box center coordinates. By training the prediction function $g_\eta$ on the available dataset $\mathcal{D}_2$, we aim to accurately identify the transmitter's location within the wireless environment based on the extracted semantic information and the  receive power vector from the ULA that has the user in its field of view.
	
	To learn $g_\eta$, we use a two-layered fully connected neural network with $512$ nodes in each layer. The obtained $\mathbf{b}_{\text{Tx}}$ from $g_\eta$ is not intended to be the final prediction but rather only an initial estimate based solely on wireless data. The idea here is to utilize this initial estimate together with the semantic information at that time instant to identify the bounding box and mask of the object responsible for the received signal. The bounding box of the transmitter is identified by locating the bounding box in $\mathbf{X}_{\text{BBox}}[\tau-r+1]$  and $\mathbf{X}_{\text{B-Mask}}[\tau-r+1]$ whose center coordinate is closest to $\mathbf{\hat{b}}_\text{Tx}[\tau -r +1]$.   Furthermore, we determine the transmitter's mask by identifying the group of pixels within the transmitter's bounding box. We assume that a well-trained prediction function $g_\eta$ can approximate the center coordinates near the actual values, and hence the Euclidean distance-based metric can effectively identify the transmitter. It is worth noting here that this work does not account for the no-transmitter situation, i.e., we assume that there is a transmitter present in the wireless environment at each time step $t$. The next step involves tracking the bounding box and mask of the transmitter for the next $r-1$ samples. 
	
	\subsubsection{Object Association Based Tracking} In the previous step, we successfully identified the transmitter in the scene based on the first image sample (in a sequence of $r$ images), and the receive power vector corresponding to that image sample. However, to predict the current optimal beam index, tracking the transmitter's location throughout the remaining $r-1$ samples in the sequence is essential. This tracking process allows us to capture the transmitter's movements and ensure accurate beam prediction. This section presents two distinct approaches for transmitter tracking for the different environment semantics: (i) Bbox-based object tracking and (ii) mask-based object tracking.
	
	\textbf{(i) Bbox-based Object Tracking:} Numerous state-of-the-art algorithms \cite{8239743, 9578284, 9305423} have been proposed in the field of multiple object tracking (MOT), which continues to be an active research area. However, considering the emphasis of this work on V2I communication, primarily involving mobile vehicles (where the users typically move in easy-to-predict mobility patterns), a simple Euclidean distance-based object association algorithm is adopted \cite{charan2023user}. This algorithm determines the transmitter in the next sample by finding the bounding box in $\mathbf{X}_{\text{BBox}}$ (of the following sample) with the closest center coordinate to the bounding box in the current sample as shown in Fig. \ref{fig:trans_iden}. The key underlying idea is that, for two consecutive image samples, the distance between the center coordinates of the bounding box will be the smallest for the same object compared to other objects in the scene.
	
	\textbf{(ii) Mask-based Object Tracking:} To facilitate object association-based tracking using masks, the median color value of mobile vehicles can be utilized. Using binary masks, we extract the color information of all the detected vehicles at the distributed nodes. This is achieved by performing a Hadamard product between the binary mask and the RGB image, followed by calculating the mean value of the pixels where the binary mask contains a $1$. The color information, binary masks, and bounding boxes are then utilized to enhance the object association-based tracking accuracy following the transmitter identification step. In particular, this allows us to filter out the vehicles whose color does not match that of the transmitter identified in the first sample. Let $\rho \in \mathbb{R}^3$ denote the median RGB color value of a probable candidate. Let $\rho_{\text{Tx}}$ and $\rho_z$ further represent the median RGB color values of the transmitter and the $z^{th}$ potential user in the mask, respectively. The potential user is considered a candidate for subsequent object association if the following criterion is satisfied
	\begin{equation}
		\| \rho_{\text{Tx}} - \rho_z \|_F \leq \epsilon,
	\end{equation}
	where $\epsilon$ is a tunable threshold. This approach helps refine the object association process based on color similarity, enhancing the accuracy of the tracking algorithm. The decision of which candidate will be retained in the list of potential users depends on the choice of  $\epsilon$, which we have kept as 20. In the context of this paper, we refer to this filtering step that utilizes color information as ``semantic-aided filtering". To identify the transmitter's mask in the subsequent sample, we select the mask of the vehicle with the shortest distance to the transmitter's mask in the previous frame as the nearest neighbor. Consequently, this selected mask is designated as the transmitter's mask in the subsequent frame. By incorporating color similarity as a refining criterion in the object association process, we enhance the accuracy of the tracking algorithm.
	
	\begin{figure*}
		\centering
		\subfigure[RNN model taking bounding box as input]{\includegraphics[width=0.475\linewidth]{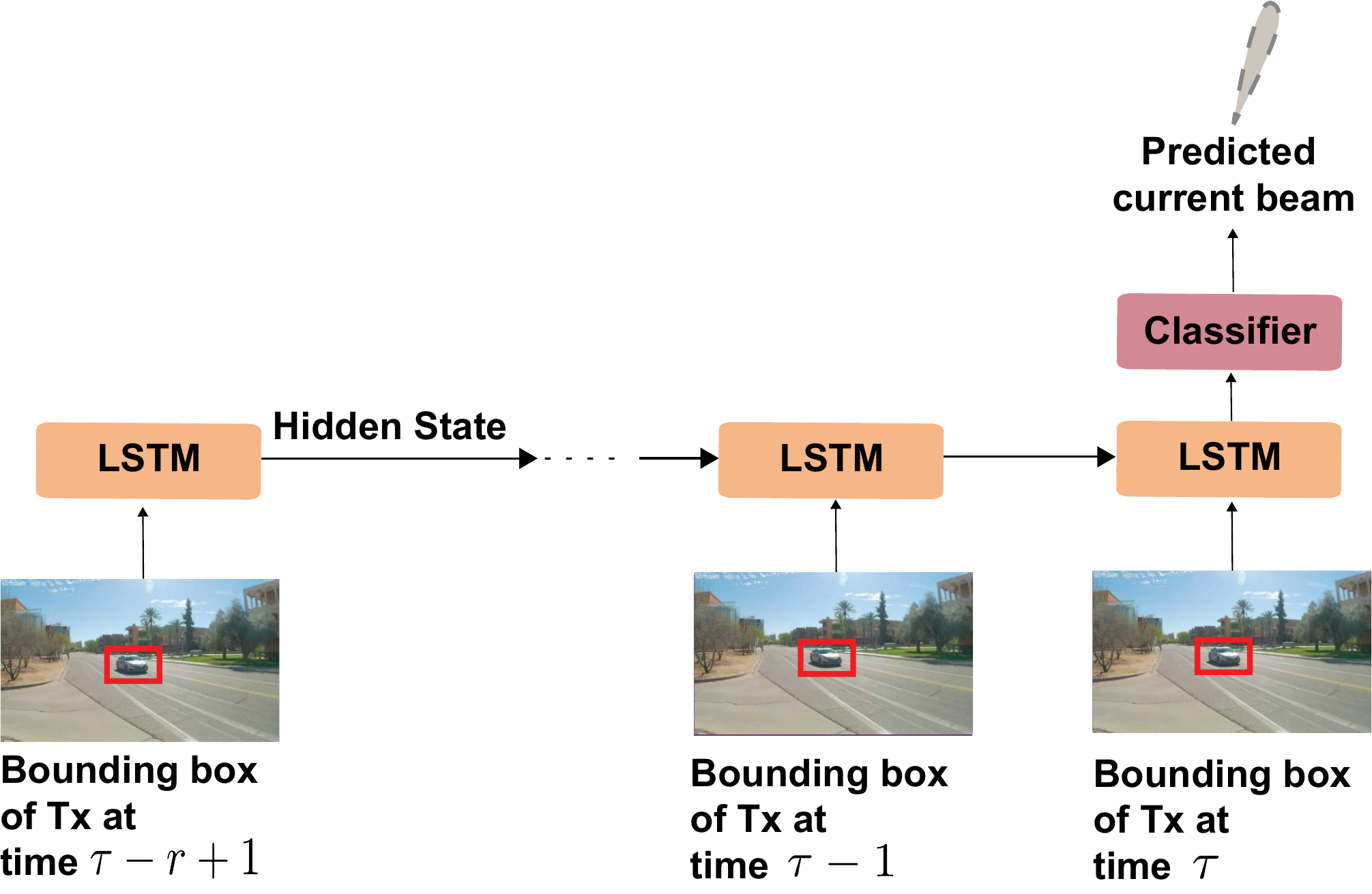}
			\label{fig:LSTM_subfigure1}}
		\hfill
		\subfigure[RNN model taking mask as input]{\includegraphics[width=0.475\linewidth]{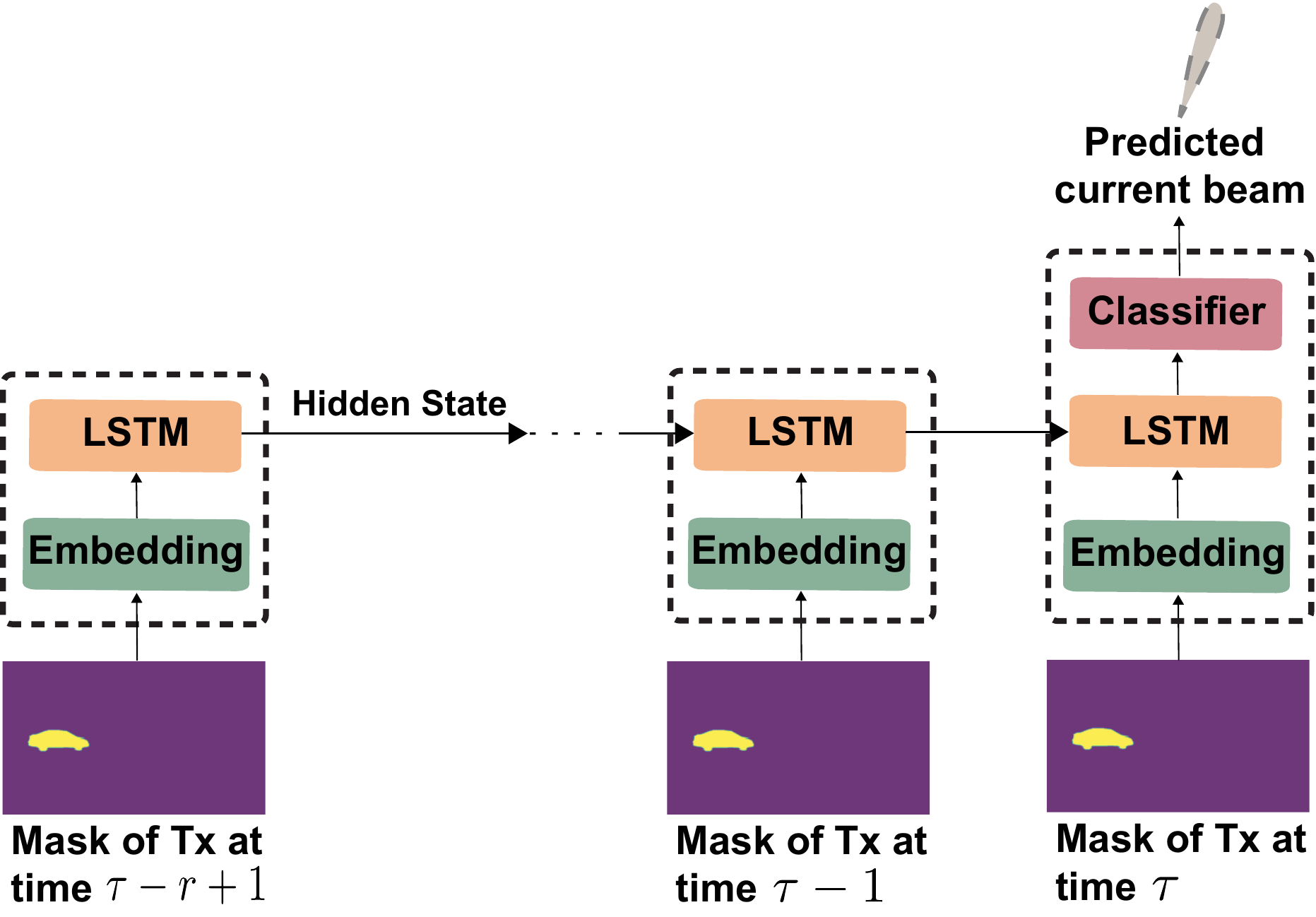}
			\label{fig:LSTM_subfigure2}}
		\caption{The figures show the proposed RNN models for beam prediction. The first RNN model, shown in (a), takes the bounding boxes of the transmitter as input. Each unit consists of an LSTM block and a classifier block. The RNN model shown in (b) takes masks of the transmitter as input. Each unit consists of an embedding block, an LSTM block, and a classifier block.}
		\label{fig:LSTM_main_figure}
	\end{figure*}

	\subsection{Stage 3: Beam Prediction}\label{Sec:Beam_Prediction}
	This section introduces the final step of our proposed solution, which aims to predict the optimal beam index for the transmitter. The goal is to use the sequence of bounding-box coordinates or image masks obtained from the previous object association-based tracking to make this prediction. However, since we are interested in predicting the current optimal beam index rather than future ones, it may be sufficient to use the available semantics for the current time step only. In order to address this, we propose two approaches: (i) Single instance-based beam prediction and (ii) Sequence-based beam prediction. In the single instance-based approach, we use the bounding box or mask at the current time step $t$ to predict the optimal beams. For the sequence-based approach, we utilize the sequence of $r$ available environment semantics to make the prediction. The problem formulation differences require distinct approaches. Next, we present both of these proposed solutions.
	
	\subsubsection{Single Instance-based Beam Prediction} Due to the distinct nature of the environment semantics (bounding box and image mask), each requires a specific approach for predicting the optimal beam index. We present both solutions, highlighting their effectiveness in utilizing the corresponding environment semantic for accurate beam prediction.
	
	\begin{enumerate}
		\item \textbf{Bounding Box-based Beam Prediction:} This baseline model (mapping function) takes the user's bounding box at the current time instant $t$ as input and predicts the corresponding beam index. Mathematically, we can express this as 
		\begin{equation}
			\omega: \mathbf{x}_\text{bbox}[t] \rightarrow \hat{\mathbf{f}}[t], 
		\end{equation}
		
		where $\omega$ represents the mapping function and $\mathbf{x}_\text{bbox}[t] \in \mathbb{R}^{2 \times 1} $ represents the center coordinate of the transmitter vehicle's bounding box at time $t$. This mapping function takes the form of a two-layered fully connected neural network with 512 neurons in each layer as our baseline model. Fully connected neural networks (FCNNs) excel at handling structured data by leveraging the network weights to capture the relationships among input elements. Additionally, FCNNs establish dense connections between adjacent layers, enabling them to learn intricate associations between input elements. The FCNN model receives the bounding box coordinates as the input and is  trained on the labeled dataset to predict the optimal beam index.
		
		\item \textbf{Mask-based Beam Prediction:} In this step, similar to bounding box-based beam prediction, we utilize another mapping function that takes the transmitter vehicle's mask at the current time instant $t$ as input and predicts the corresponding beam index as follows
		\begin{equation}
			\beta: \mathbf{x}_\text{mask}[t] \rightarrow \hat{\mathbf{f}}[t], 
		\end{equation}
		where $\beta$ represents the mapping function for this task and $\mathbf{x}_\text{mask}[t] \in \mathbb{R}^{\hat{W} \times \hat{H}}$  represents the transmitter vehicle's mask at time $t$. We note that convolutional neural networks (CNNs) have demonstrated superior performance and robustness in leveraging spatial relationships among neighboring pixels in image data. Therefore, the mapping function $\beta$ for this baseline model of mask-based beam prediction takes the form of the simple CNN model of LeNet \cite{lecun1998gradient}. The LeNet model consists of two convolutional layers followed by two fully connected layers. Taking the mask as input, the LeNet model is trained to accurately predict the optimal beam indices.
	\end{enumerate}

	\begin{figure*}
		\centering
		\includegraphics[width=1\linewidth]{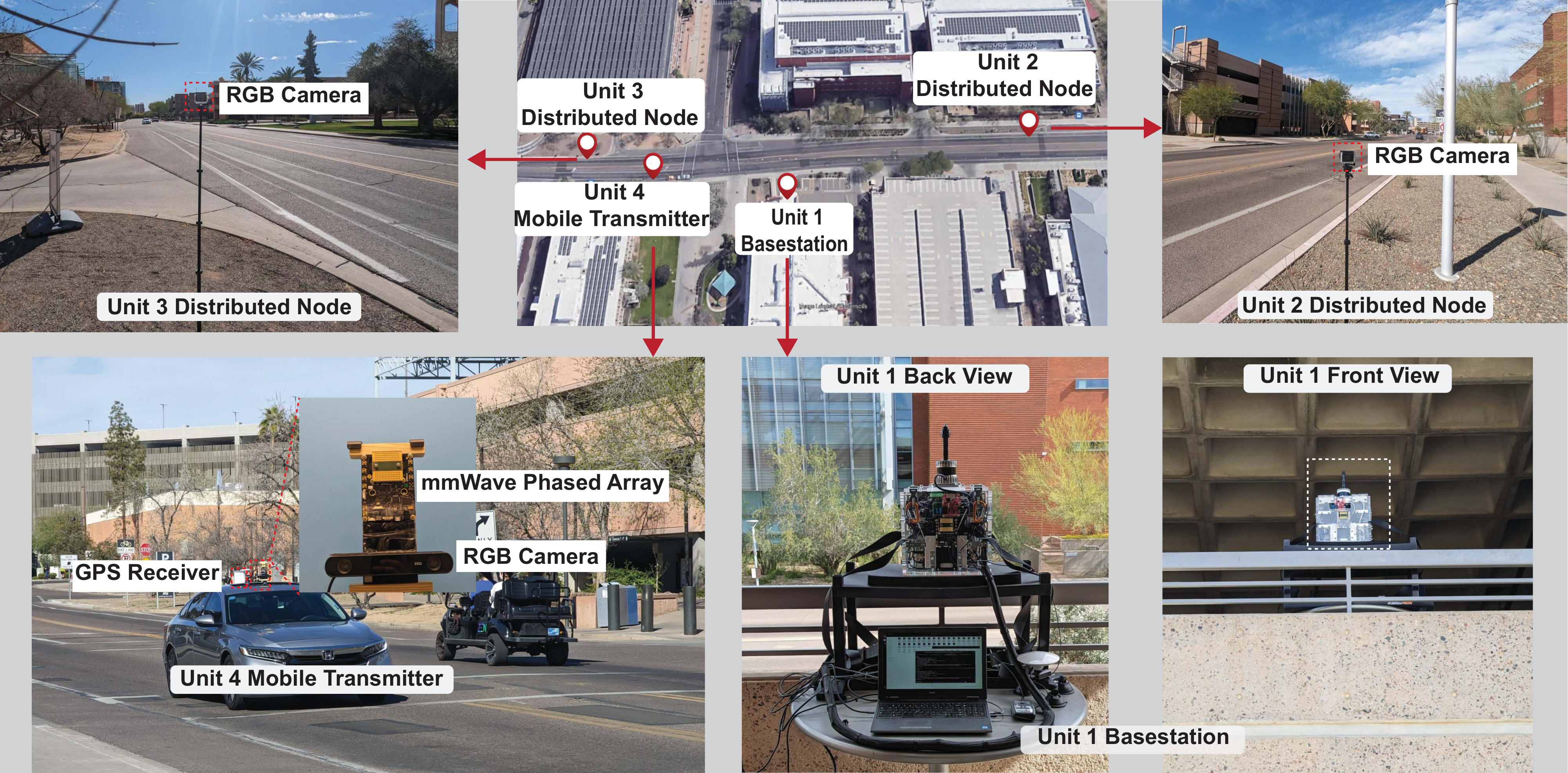}
		\caption{The figure illustrates the testbed setup for the DeepSense 6G AI-ready dataset used in our experiments. It consists of a stationary unit (unit 1), acting as the basestation, a mobile unit (unit 4), acting as the transmitter, and two distributed nodes (unit 2 and unit 3). }
		\label{fig:testbed}
	\end{figure*}

	\subsubsection{Sequence-based Beam Prediction} We use a recurrent neural network (RNN) \cite{cho2014learning, greff2016lstm} which processes a sequence of semantic representations of the transmitter and predicts the optimal beam index as the output. We chose the RNN architecture for two reasons. First, RNNs have achieved good accuracy in various sequential modeling tasks, such as natural language processing and speech recognition, due to their ability to extract crucial information from previous sensory data. This approach allows the model to capture the temporal dependencies and patterns in the semantic information, enabling accurate beam prediction. By considering the historical sequence of transmitter representations, the RNN model can effectively learn the correlations between the semantic information and the optimal beam selection. Second, as compared to other neural network architectures like Transformers \cite{vaswani2017attention}, RNNs offer advantages in terms of computational complexity and inference time. In this section, we present the two different solutions developed to target the different semantic modalities.
	
	\begin{enumerate}
		\item \textbf{Bounding Box-based Beam Prediction:} For sequence-based beam prediction with bounding box as inputs, we utilize a mapping function that takes a sequence of the transmitter vehicle's bounding boxes over $r$ consecutive time stamps and predicts the corresponding beam index at the last time step. Mathematically, we can express this as 
		\begin{equation}
			\gamma: \left\{\mathbf{x}_\text{bbox}[t] \right\}_{t = \tau-r+1}^{t= \tau}\rightarrow \hat{\mathbf{f}}[\tau], 
		\end{equation}
		where $\gamma$ represents the mapping function for bounding box sequence-based beam prediction. The mapping function $\gamma$ takes the shape of a RNN model. In Fig. \ref{fig:LSTM_subfigure1}, we present the block diagram of the proposed RNN model for beam prediction using bounding boxes as inputs. This model comprises $r$ repeated blocks, each consisting of a Long Short-Term Memory (LSTM) unit. The bounding box vectors are directly fed into the LSTM block. The hidden state of the LSTM is initialized with all-zero vectors. The final component of the model is the classifier, which utilizes the cross-entropy activation function. The output of the classifier is a score vector. The beam index with the highest score is the predicted optimal beam.
		
		\item \textbf{Mask-based Beam Prediction:} 
		In this step, similar to bounding box sequence based-beam prediction, we utilize a mapping function that takes a sequence of the transmitter vehicle masks over $r$ consecutive time stamps and predicts the beam index at the last time step. We can formally express this as
		\begin{equation}
			\psi: \left\{\mathbf{x}_\text{mask}[t] \right\}_{t = \tau-r+1}^{t= \tau}\rightarrow \hat{\mathbf{f}}[\tau], 
		\end{equation}
		where $\psi$ represents the mapping function for this task. This mapping function again takes the shape of a RNN as shown in Fig.  \ref{fig:LSTM_subfigure2}. This model also consists of $r$ repeated blocks, each comprising an LSTM unit. However, due to the structural differences between masks and bounding box vectors in terms of semantic representation, an additional embedding block is included in this model. The embedding block utilizes the LeNet model, consisting of two convolutional layers and two fully connected layers. The final output layer of the LeNet model is removed, and the output from the prefinal layer is used as input to the LSTM model. It transforms the high-dimensional semantic mask $\mathbf{x}_{\text{mask}}[t]$ into a low-dimensional embedded vector $x[t] \in \mathbb{R}^{\nu \times 1}$, where $\nu$ represents the hidden state size of the LSTM. The LSTM hidden state is initialized with all-zero vectors. The remaining components, including the LSTM blocks and the classifier with the cross-entropy activation function, are similar to the bounding box-based model. The model predicts the optimal beam index based on the scores obtained from the classifier.
		
	\end{enumerate}
	
	In conclusion, we propose two different models that are designed to effectively capture the relevant information from the semantic representations (bounding box and mask) and predict the optimal beams accurately.

	\subsection{Baseline Solutions}\label{Sec:baseline_sol}
	In order to evaluate the accuracy of our proposed transmitter identification solution and the subsequent object association step, it is important to have ground-truth bounding box center coordinates of the transmitter for all the samples in the dataset. However, since such detailed information is not available, we propose an alternative approach that leverages the transmitter's GPS position to aid in the identification process. The reason for selecting position as an additional input is that the dataset used in this study provides highly accurate position data. By incorporating this precise position information into the prediction model, we can expect a significant improvement in the accuracy of transmitter identification. The position data offers more reliable and granular information about the location of the transmitter, enabling the model to make more precise predictions. In the position-aided transmitter identification approach, a machine-learning model predicts the center coordinate of the transmitter's bounding box based on its GPS position. The network architecture for this position-aided identification is identical to the receive power vector-aided model, consisting of a two-layered fully connected neural network with 512 neurons in each layer. However, unlike the previous solution, which involved identifying the transmitter only at the time step of the first sample in the sequence and then tracking it for the remaining frames, the current approach performs transmitter identification at every time step of the sequence. The results obtained from the position-aided transmitter identification serve as a baseline for evaluating the performance of the transmitter identification using the receive power vector and the subsequent object association-based transmitter tracking.

	%####################################################################################
	%####################################################################################
	%####################################################################################

	\section{Testbed Description and AI-Ready Dataset}\label{sec:testbed}
	To assess the effectiveness of our proposed distributed sensing-aided beam prediction solution, we employ the DeepSense 6G dataset \cite{DeepSense}. DeepSense 6G is a comprehensive real-world dataset specifically designed for sensing-aided wireless communication applications. It encompasses diverse multi-modal data, including vision, mmWave wireless communication, GPS data, LiDAR, and radar, all collected in real-world wireless environment. In this section, we provide an overview of scenario $40$ selected from the DeepSense 6G dataset, and subsequently analyze the AI-ready dataset used to evaluate the performance of our proposed solution.
	
	\subsection{DeepSense 6G Testbed}\label{sec:deepsense_testbed}
	The study adopts scenario $40$ of the DeepSense 6G dataset specifically designed to study distributed sensing-aided communication in a multi-user scenario. The hardware testbed and the locations for collecting this data are shown in Fig. \ref{fig:testbed}. The DeepSense testbed 7 is utilized for this data collection. It consists of  (i) three stationary units, one acting as the basestation and the other two acting as the distributed nodes, and (ii) a mobile transmitter (vehicle). All the stationary units, namely the basestation (unit 1), the first distributed node (unit 2), and the second distributed node (unit 3), are equipped with an RGB camera. The basestation further adopts three 16-element ($M$=16) 60 GHz-band phased arrays, and it receives the transmitted signal using an over-sampled codebook of 64 pre-defined beams($Q=64$). In this data collection scenario, the mobile unit (unit 4) is a vehicle equipped with a mmWave transmitter and GPS antenna/receiver. The transmitter consists of a quasi-omni antenna constantly transmitting (omnidirectional) at the 60 GHz band. For more information regarding the data collection setup and testbed, please refer to \cite{DeepSense}.
	
	\begin{figure}
		\centering
		\includegraphics[width=1\linewidth]{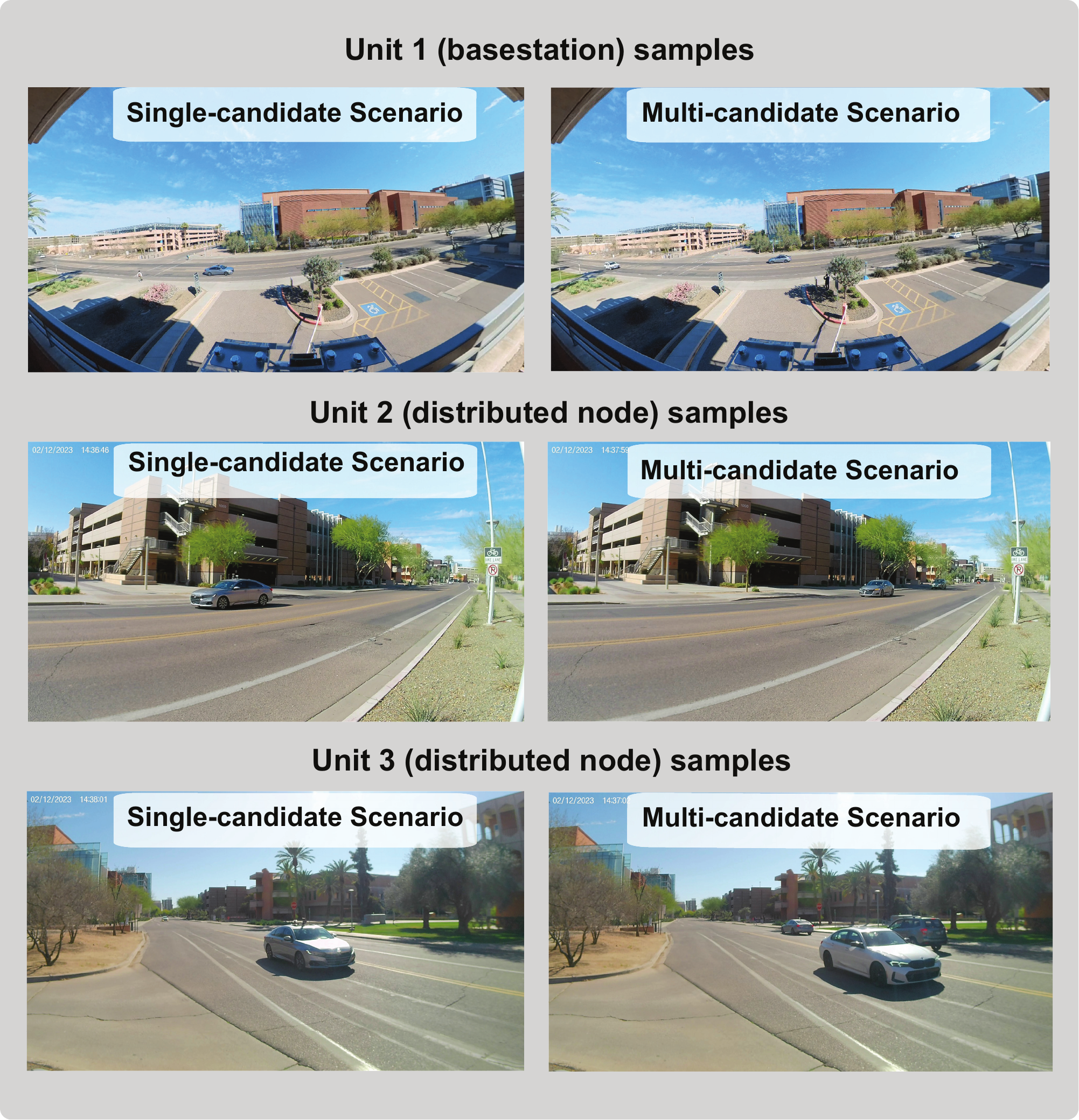}
		\caption{The figure presents the RGB image samples captured at the basestation (unit 1) and the distributed nodes (units 2 and 3), illustrating both single-candidate and multi-candidate scenarios.}
		\label{fig:samples}
	\end{figure}
	
	\subsection{Deepsense 6G AI-Ready Dataset}\label{sec:develop_dataset}
	The evaluation of the proposed distributed sensing-aided beam prediction solution necessitates real-world data obtained from a wireless environment featuring a moving vehicle as the mmWave transmitter. As such, we utilize scenario 41 of the DeepSense 6G dataset. This dataset is collected at McAllister Ave., Tempe, during the daytime. Throughout the data collection process, the road was actively utilized by other vehicles, pedestrians, and cyclists. The raw dataset includes RGB images from both the basestation (unit 1) and the distributed nodes (unit 2 and unit 3), receive power vectors from the three ULAs, and the user's GPS position. Fig. \ref{fig:samples} shows the sample dataset images from each unit. To prepare the AI-ready dataset for our experiments, we process the RGB images from unit 2 and unit 3 using a sliding window of size $r=5$, generating time-series sequences of RGB images for each unit. The AI-ready dataset comprises these processed RGB image sequences, along with the receive power at the initial time step, $\mathbf{p}[\tau-r+1]$, and the optimal beam index $\mathbf f^\star$ at the last time step of each sequence. Furthermore, it also incorporates the transmitter's GPS position at every time instant. Only the sequences where the transmitter car is present in the camera's field of view are retained in the AI-ready dataset. There are 2991 and 5476 such image sequences for unit 2 and unit 3, respectively, which are further split into training, validation, and testing categories with a ratio of 70:20:10. 
	
	For the transmitter identification models, we construct separate datasets for each node. The dataset for the position-aided transmitter identification models consists of pairs of GPS positions and the corresponding center coordinates of the transmitter's bounding box. Similarly, the dataset for the receive power-aided transmitter identification models includes pairs of receive power vectors and their corresponding center coordinates. Rather than manually labeling the center coordinates, we select the samples where only the transmitter car is present in the scene. We have 343 and 1124 such samples for unit 2 and unit 3, respectively. The bounding box center coordinates of the transmitter vehicle in these images and their corresponding positions and receive power vectors form the dataset for the transmitter identification models. 
	
	%####################################################################################
	%####################################################################################
	%####################################################################################
	
	\section{Performance Evaluation} \label{sec:perf_eval}
	This section focuses on evaluating the performance of the proposed distributed sensing-aided beam prediction solution. In Section \ref{sec:Exp_set}, we provide a description of the experimental setup utilized in this work. We then analyze the results of the proposed solution in Section \ref{sec:Num_results}.
	
	\subsection{Experimental Setup}\label{sec:Exp_set}
	We first outline the neural network training parameters of the machine learning models adopted in this work. Next, we discuss the evaluation metrics which we utilize to assess the performance of different stages of the proposed solution.
	
	\begin{table*}[!t]
		\caption{Beam Prediction: Design and Training Hyper-parameters}
		\centering
		\setlength{\tabcolsep}{5pt}
		\renewcommand{\arraystretch}{1.2}
		\begin{tabular}{@{}l|cc|cc|cc@{}}
			\midrule \midrule
			\textbf{ML Model} & Mask-LSTM & BBox-LSTM & Mask-LeNet & Bbox-FCNN & Position-Aided FCNN  & Receive Power-Aided FCNN\\
			\textbf{Batch Size} & 5 & 8 & 5 & 8 & 50 & 50 \\
			\textbf{Learning Rate} & $1 \times 10^{-3}$ & $1 \times 10^{-2}$ & $1 \times 10^{-3}$ & $1 \times 10^{-2}$ & $1 \times 10^{-2}$ & $1 \times 10^{-2}$ \\
			\textbf{Learning Rate Decay} & - & epoch 20 & - & epochs 20 & epochs 30 and 70 & epochs 30 and 70 \\
			\textbf{LR Reduction Factor} & - & 0.1 & - & 0.1 & 0.1 & 0.1 \\
			%\textbf{Dropout} & 0.3 & 0.3 & 0.3 \\
			\textbf{Total Training Epochs} & 50 & 50 & 50 & 50 & 100 & 100 \\ 
			\bottomrule \bottomrule
		\end{tabular}
		\label{tab_beam_pred_train_params}
	\end{table*}
	
	\textbf{Network Training:}  As described in Section \ref{sec:propsol}, the proposed distributed sensing-aided beam prediction solution consists of three steps: 1) environment semantics extraction 2) transmitter identification and tracking  and 3) beam prediction. In the transmitter identification and tracking stage, we use a two-layered fully connected neural network with 512 neurons in each layer to predict the center coordinates of the transmitter's bounding box within the image. For the beam prediction stage, we employ distinct LSTM models for bounding box-based beam prediction and mask-based beam prediction, as elaborated in Section \ref{Sec:Beam_Prediction}. In the case of bounding-box based beam prediction, we employ a baseline model consisting of a two-layered FCNN with 512 neurons in each layer.  For mask-based beam prediction, we evaluate the LSTM model for it against the LeNet CNN model. In the beam prediction classification task, the LSTM models and their respective baselines are trained using cross entropy loss. On the other hand, the receive power-aided transmitter identification FCNN and its corresponding baseline FCNN are trained using mean squared error loss. In the transmitter identification regression task, both the FCNNs, one taking receive power vector as input and the other taking position as input, are trained using mean squared error  loss. We use Adam optimizer to train all the aforementioned models. These models are trained on the AI-ready dataset discussed in Section  \ref{sec:testbed}. The detailed hyperparameters used to fine-tune each model are presented in Table \ref{tab_beam_pred_train_params}.

	\textbf{Evaluation Metrics:} The key evaluation metric used to assess the proposed beam prediction solution is the top-$k$ accuracy, which measures the percentage of test samples where the optimal ground-truth beam falls within the top-$k$ predicted beams. In this work, we present the top-1, top-2 and top-3 accuracies to comprehensively evaluate the performance of the beam prediction stage. 
	
	Furthermore, we use the metric of association accuracy to evaluate the performance of tracking the transmitter. Note that the association accuracy varies across different frames. The association accuracy for a specific frame is defined as the percentage of samples for which the transmitter predicted by the receive power-aided FCNN matches with the one predicted by the position-aided FCNN. Note that this calculation assumes that the transmitter was correctly identified in the first frame i.e.  both the receive power-based and position-based FCNNs identify the same object as  transmitter in the initial frame. Moreover, in computing association accuracy, it is important to note that we do not include the sequences where the difference between the predicted center coordinates by the position-aided FCNN and the center coordinate of the closest bounding box in $\mathbf{X}_{\text{BBox}}$ and $\mathbf{X}_{\text{B-Mask}}$ exceeds a specified threshold.
	\begin{figure}[t]
		\centering
		\includegraphics[width=1\linewidth]{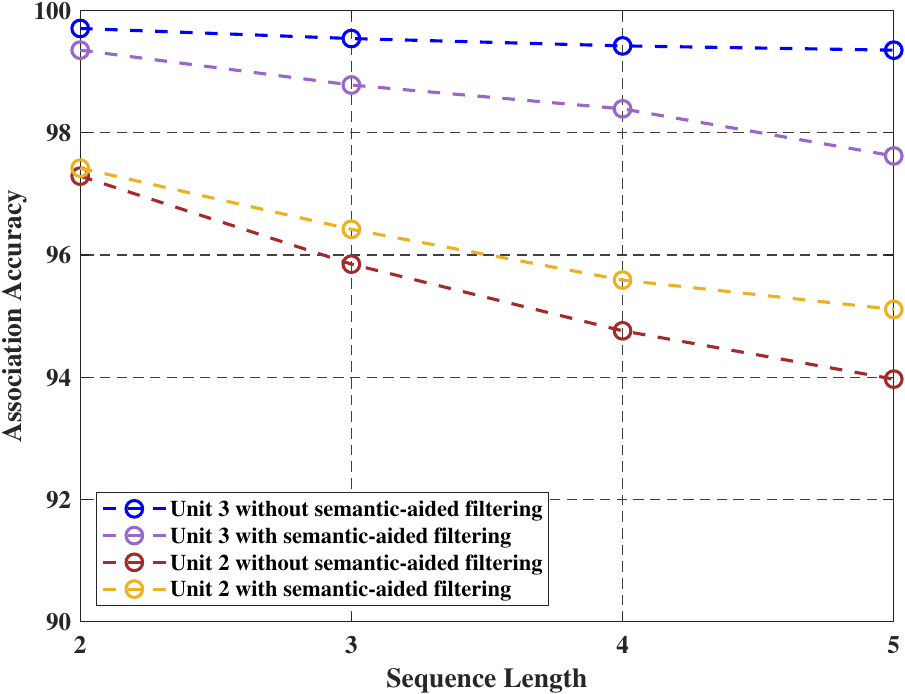}
		\caption{The figure shows the variation of the association accuracy with sequence length for both units 2 and 3 with and without semantic-aided filtering. We observe that semantic-aided filtering has contrasting effects on the association accuracy from both units 2 and 3 across all sequence lengths. Semantic-aided filtering increases the association accuracy for unit 2 while decreasing it for unit 3.}
		\label{fig:assoc_acc_vs_seq_len}
	\end{figure}
	
	\subsection{Numerical Results}\label{sec:Num_results}
	This section presents a detailed evaluation of the results from the proposed solution. First we evaluate the accuracy of object association within the sequence. We then analyze the beam prediction accuracy for the two machine learning solutions outlined in Section \ref{sec:propsol}. The first solution utilizes all the samples in the sequence for beam prediction, while the second solution only considers the last sample in the sequence. In addition, we explore the impact of the distance between the mobile transmitter and the distributed node on beam prediction accuracy. We also examine how the beam prediction accuracy is affected by the increasing the  number of mobile objects present in the wireless scene. In doing so, we aim to address the following questions.

	\textbf{ To what extent can the proposed object association based-method accurately track the transmitter?} 
	
	As presented in Section~\ref{subsec:txid_track}, after the transmitter is identified in the first frame of the sequence, the proposed solution involves tracking the transmitter for the next $r-1$ samples. Fig. \ref{fig:assoc_acc_vs_seq_len} shows how the association accuracy varies against the sequence length with and without semantic-aided filtering for both unit 2 and unit 3. We observe that the association accuracy for unit 3 decreases only marginally as the sequence length increases and remains above 99\% for the whole length of the sequence. However, the association accuracy for unit 2 decreases as the sequence length increases. This may be attributed to the different environmental conditions that the cameras of each unit face. These environmental conditions can include anything from lighting conditions to traffic stoppages. We also observe that semantic-aided filtering increases the association accuracy for unit 2 but decreases it for unit 3. This decrease in association accuracy can also be attributed to the environmental conditions. For instance, if the sub-region for unit 3 consists of both shades and sunny sides, then the color information of the mobile user may change significantly.

	\begin{figure}
		\centering
		\subfigure[Top-$k$ beam prediction accuracies for unit 2]{\includegraphics[width=1\linewidth]{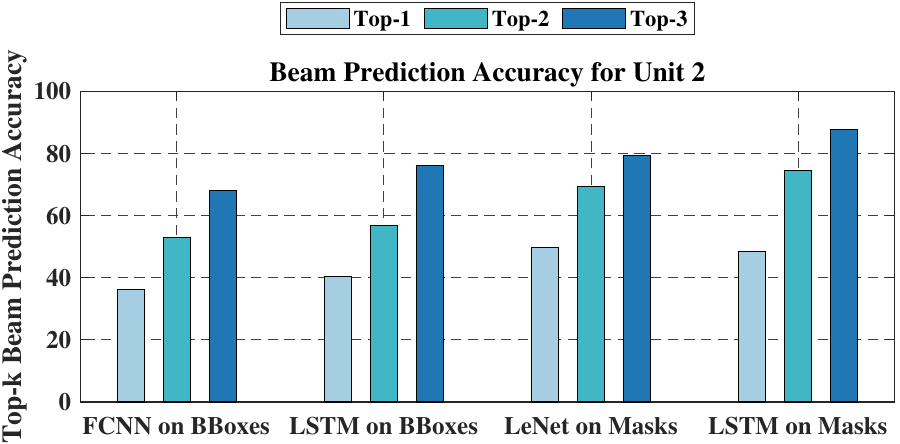}
			\label{fig:subfigure_1_fig:beam_pred_acc}}
		\hfill
		\subfigure[Top-$k$ beam prediction accuracies for unit 3]{\includegraphics[width=1\linewidth]{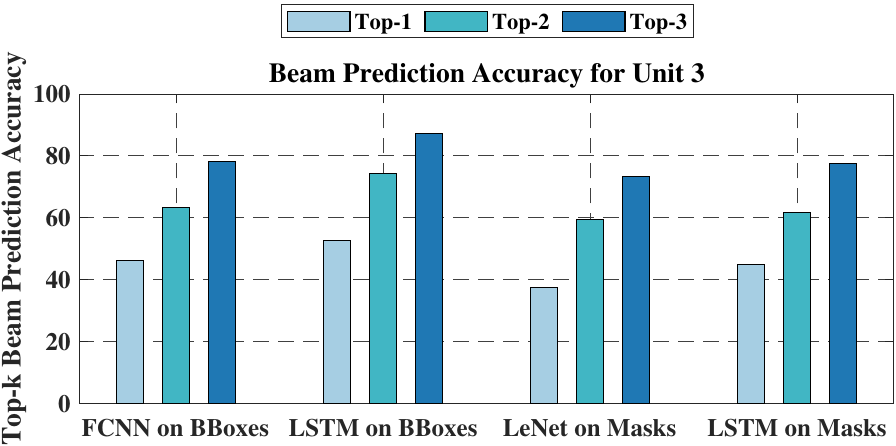}
			\label{fig:subfigure_2_fig:beam_pred_acc}}
		\caption{The figure compares the beam prediction accuracies of the proposed LSTM models for both units 2 and 3. Overall, the LSTM models, which consider a sequence of environment semantic information as input, achieve better beam prediction accuracy than the solutions that only use the last sample's semantic information as input.}
		\label{fig:beam_pred_acc}
	\end{figure}

	\begin{figure*}[t]
		\centering
		\subfigure[]{
			\includegraphics[width=0.22\linewidth]{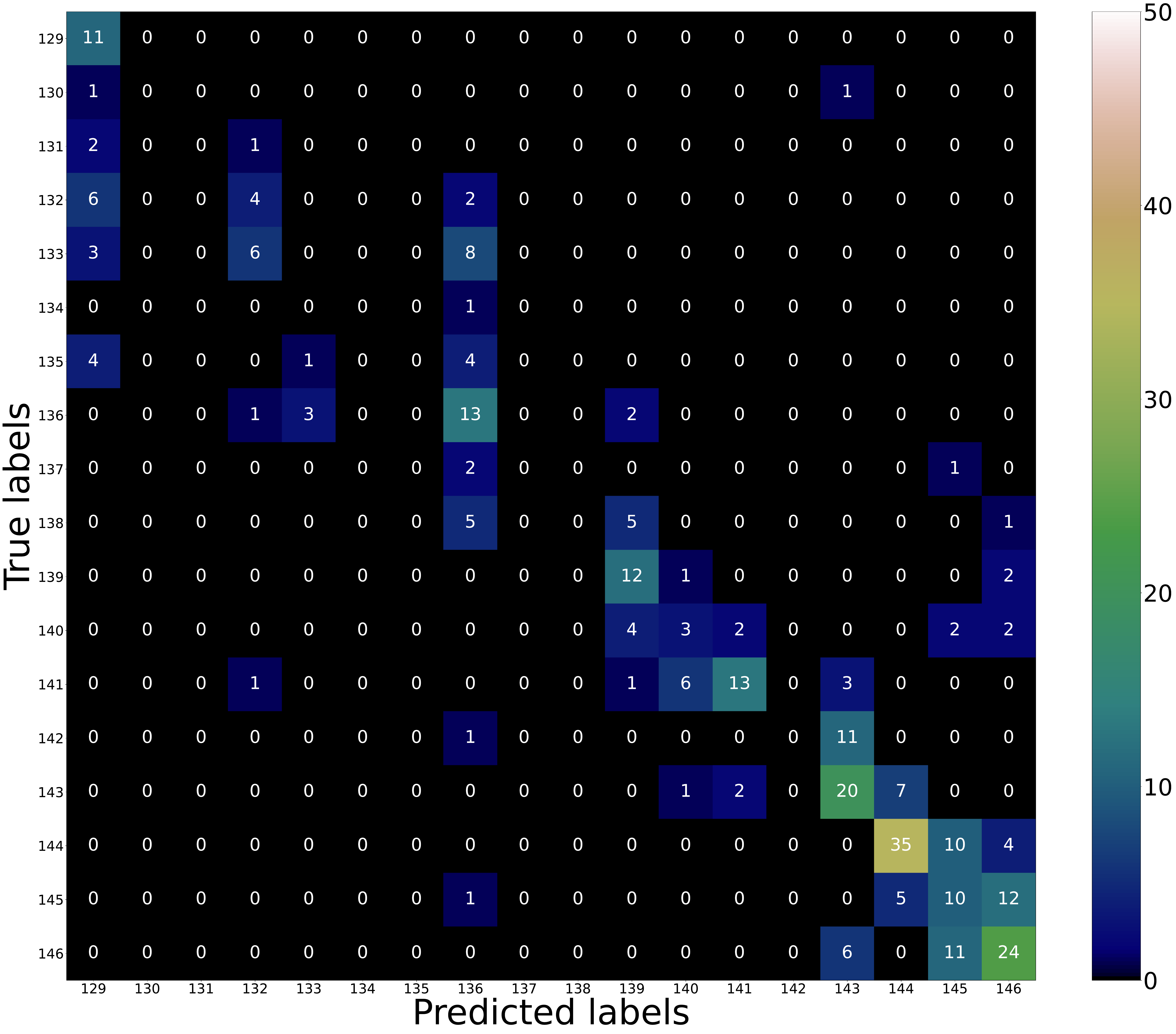}
			\label{subfig:unit2_mask_confusion_matrix}
		}
		\subfigure[]{
			\includegraphics[width=0.22\linewidth]{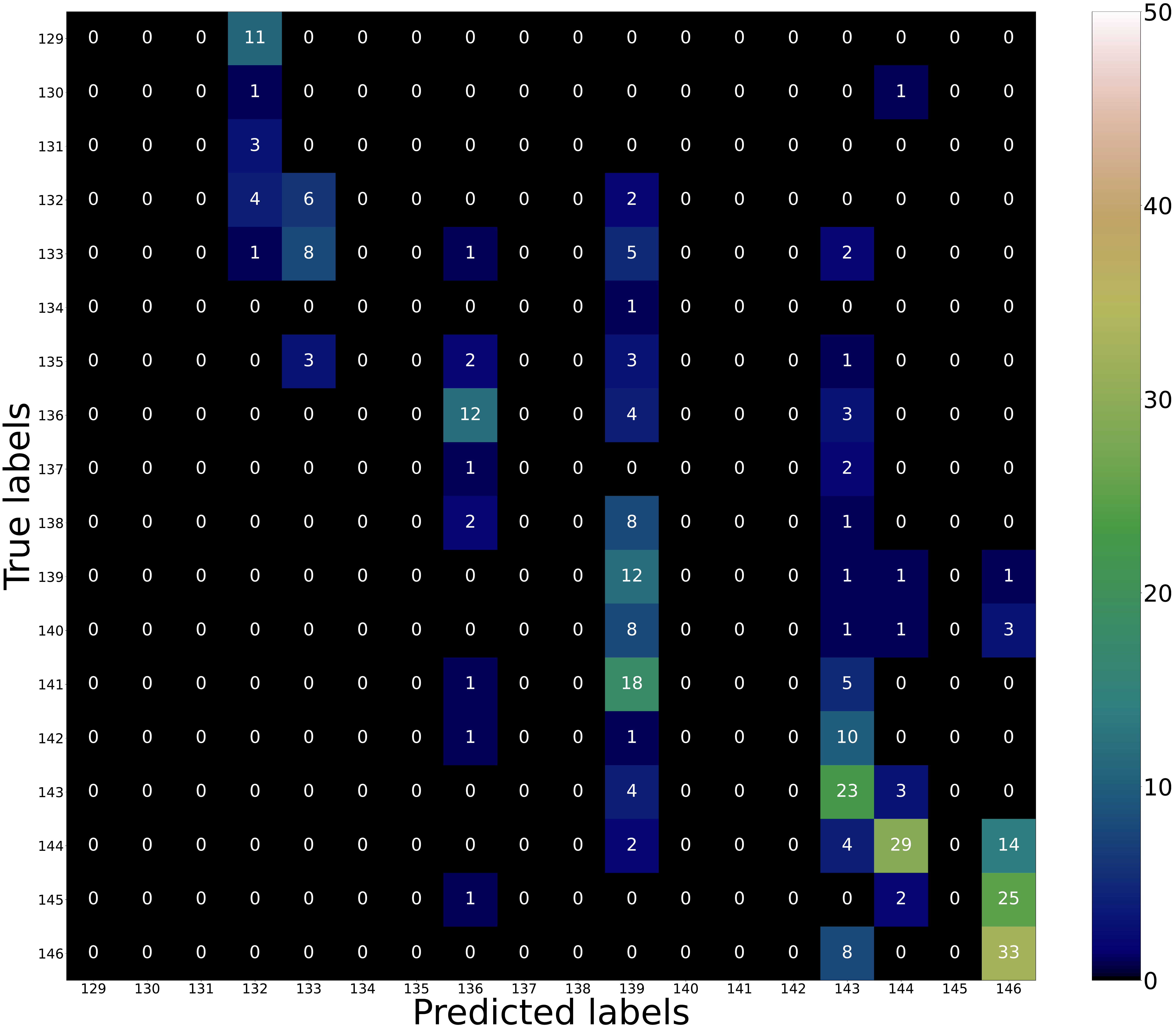}
			\label{subfig:unit2_bbox_confusion_matrix}
		}
		\subfigure[]{
			\includegraphics[width=0.22\linewidth]{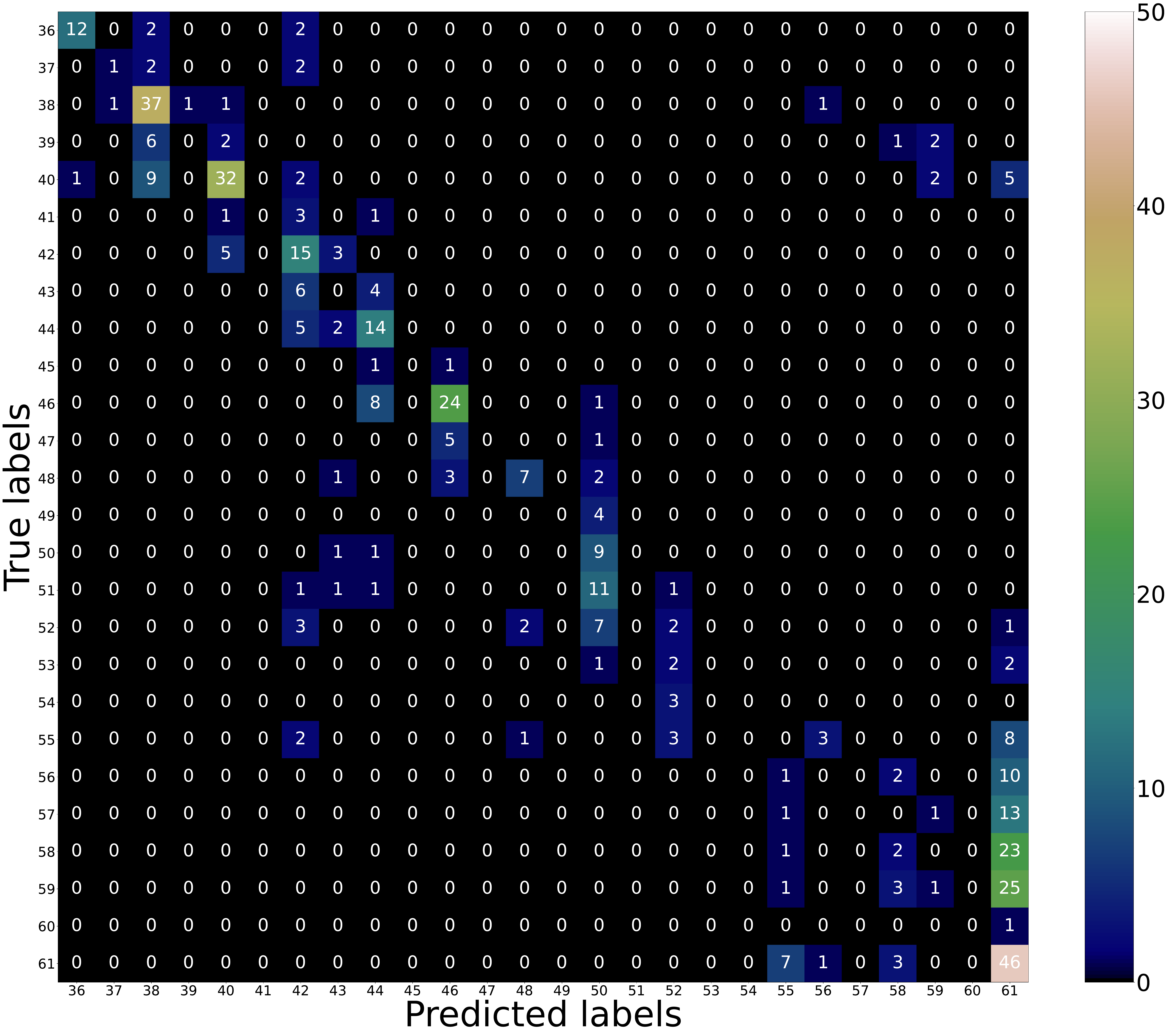}
			\label{subfig:unit3_mask_confusion_matrix}
		}
		\subfigure[]{
			\includegraphics[width=0.22\linewidth]{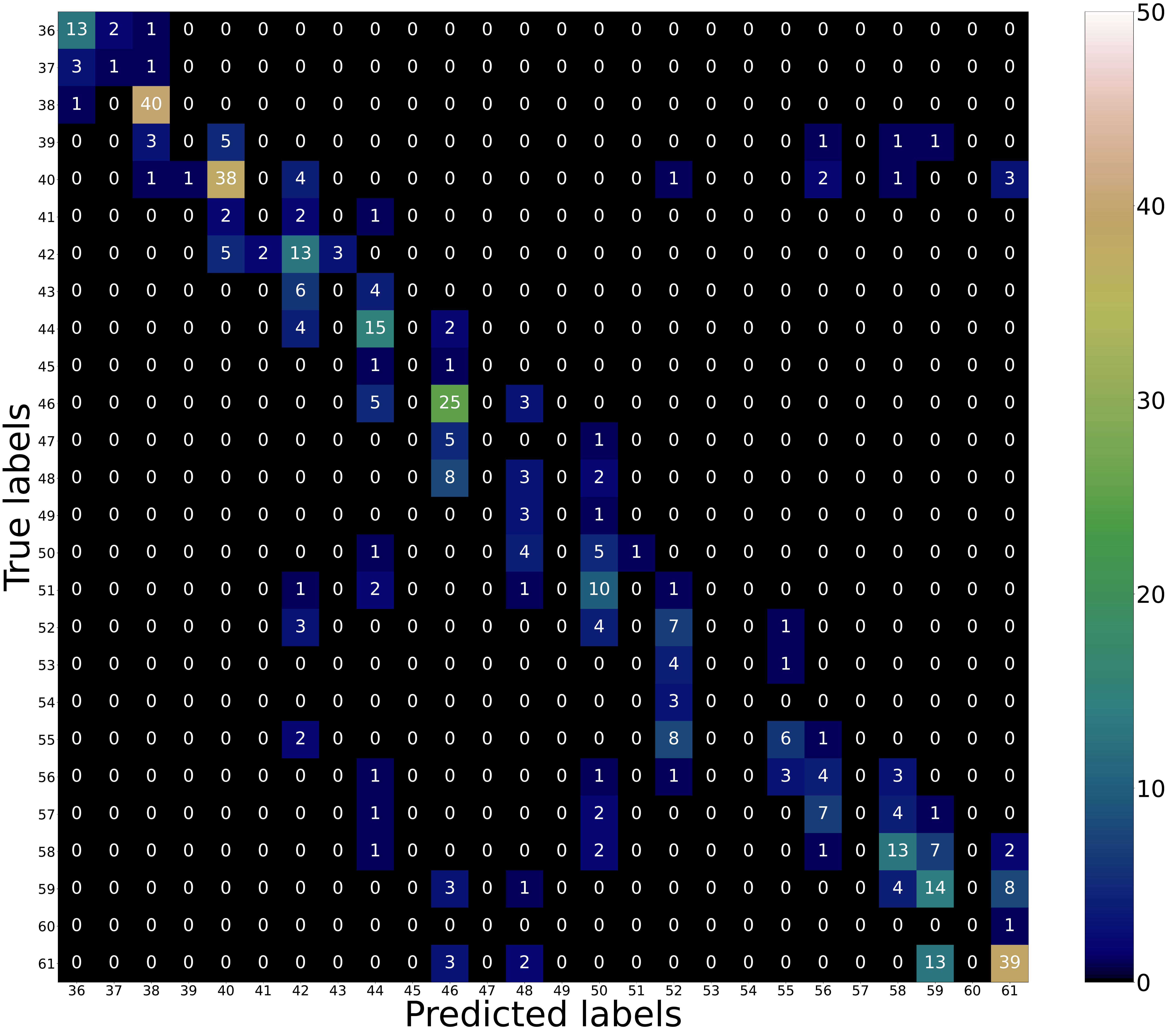}
			\label{subfig:unit3_bbox_confusion_matrix}
		}
		\caption{Fig. (a) and (b) present the confusion matrix plots for unit 2, showing the results obtained from the mask-based LSTM model and bounding box-based LSTM model, respectively. On the other hand,  Fig. (c) and (d) present the confusion matrix plots for unit 3, showing the results obtained from the mask-based LSTM model and bounding box-based LSTM model, respectively. The mask-based LSTM model for unit 2 gives more correct predictions than the bounding box-based LSTM model. For unit 3, however, the bounding box-based LSTM model gives more correct predictions than the mask-based LSTM model.}
		\label{fig_conf_matrix}
	\end{figure*}
	
	\begin{figure}
		\centering
		\includegraphics[width=1\linewidth]{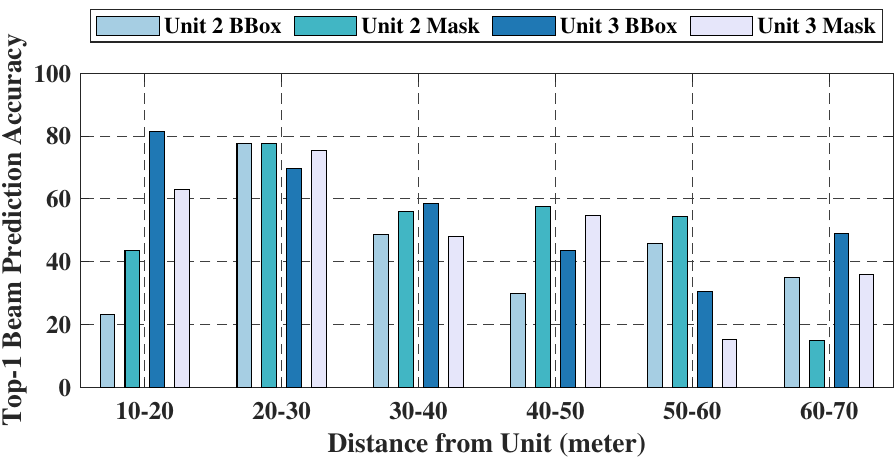}
		\caption{The figure illustrates how the top-1 beam prediction accuracies of the LSTM models change with the distance from the distributed node for both units 2 and 3. We note that in the case of both units 2 and 3, the mask-based LSTM model achieves higher beam prediction accuracy for certain distances, while the bounding box-based LSTM model performs better for other distances. This shows that certain semantic representations may be more effective in certain regions than in others. }
		\label{fig:acc_vs_dist}
	\end{figure}

	\textbf{Can the environment semantics extracted from distributed nodes be used for beam prediction at the basestation?}
	
	To answer this question, we evaluate the performance of the proposed beam prediction approaches presented in Section \ref{Sec:Beam_Prediction}. Fig. \ref{fig:subfigure_1_fig:beam_pred_acc} and \ref{fig:subfigure_2_fig:beam_pred_acc} show the top-1, top-2, and top-3 beam prediction accuracies obtained for units 2 and 3 respectively.  We observe that for both units 2 and 3, the LSTM model that takes bounding boxes as input achieves better top-1, top-2, and top-3 beam prediction accuracies than the corresponding FCNN model. On the other hand, the LSTM model that takes masks as input achieves better top-2 and top-3 accuracies for unit 2 and top-1, top-2, and top-3 accuracies for unit 3 compared to the corresponding LeNet model. The top-1 accuracy obtained by the mask-based LSTM model of unit 2 is only marginally less than that obtained by the corresponding LeNet model. The improved accuracies obtained using LSTM models can be attributed to their ability to capture better the temporal dependencies and patterns in the semantic information, enabling more accurate beam prediction. We observe that the bounding box-based LSTM model performs better for unit 3, while the mask-based LSTM model performs better for unit 2 in terms of beam prediction accuracy. This suggests that specific semantic representations may be more effective in certain regions than in others. For instance, masks can capture the user's shape and orientation, which may be more beneficial for beam prediction in certain regions than in others. Furthermore, it is essential to acknowledge the disparity in the number of training sequences between units 2 and  3. The number of training sequences for unit 2 is much less than that for unit 3. We further note that the top-3 accuracy obtained from both the bounding box-based and mask-based LSTM models is more than 75\% for both units 2 and 3. This means that using either of the proposed LSTM beam prediction model, the basestation can find the optimal beam in over 75\% of instances for both units 2 and 3, thereby significantly reducing the beam training overhead to just three in the case of exhaustive search.

	Fig. \ref{subfig:unit2_mask_confusion_matrix} and \ref{subfig:unit2_bbox_confusion_matrix} show the confusion matrix plots for unit 2, utilizing masks and bounding boxes as input, respectively. We note that for unit 2, the mask-based LSTM model gives more correct predictions than the bounding box-based LSTM model. Fig. \ref{subfig:unit3_mask_confusion_matrix} and \ref{subfig:unit3_bbox_confusion_matrix} show the confusion matrix plots for unit 3 utilizing masks and bounding boxes as input, respectively. We observe that for unit 3, the bounding box-based LSTM model gives more correct predictions than the mask-based LSTM model, as evidenced by a greater number of diagonal elements in the confusion matrix of the bounding box-based LSTM model. Furthermore, it is worth mentioning that the confusion matrices of the mask-based LSTM model for unit 2 and the bounding box-based LSTM model for unit 3 show a more pronounced concentration of elements near the diagonal.

	\begin{figure}
		\centering
		\includegraphics[width=1\linewidth]{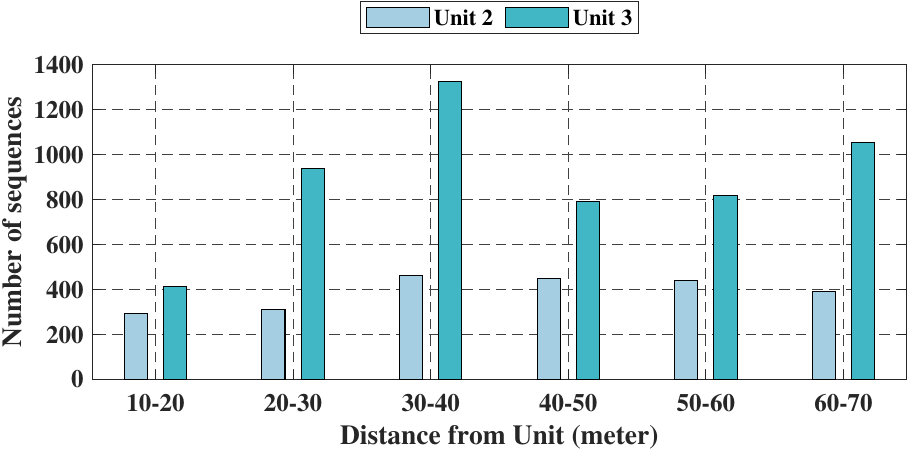}
		\caption{The figure displays the histogram showing the number of sequences falling within various distance ranges for the two distributed nodes. Unit 3 has significantly more sequences compared to unit 2. }
		\label{fig:seq_vs_dist}
	\end{figure}

	\textbf{How does the distance between the transmitter and the distributed node affect beam prediction accuracy?}
	
	Fig.  \ref{fig:acc_vs_dist} illustrates the top-1 beam prediction accuracies plotted against the distance from the distributed node for both units 2 and 3, comparing the performance of the bounding box-based LSTM model and the mask-based LSTM model. We observe that for unit 2, except for the 10-20 meter distance range, the top-1 beam prediction accuracy from the mask-based LSTM model decreases as distance increases. On the other hand, for unit 3, except for the 60-70 meter distance range, the top-1 beam prediction accuracy from the bounding box-based LSTM model decreases as distance increases. We further note that the top-1 beam prediction accuracies from the bounding box-based LSTM model of unit 2 and the mask-based LSTM model of unit 3 fluctuate across different distance ranges without showing a consistent trend. Furthermore, it is only in the distance ranges of 10-20 meters and 60-70 meters that both the bounding box-based and mask-based LSTM models of either distributed node achieve significantly better beam prediction accuracy over the other. We note that within the 10-20 meter distance range, both the LSTM models of unit 3 achieve better beam prediction accuracy as compared to unit 2. On the other hand, within the 50-60 meter distance range, both the LSTM models of unit 2 achieve better beam prediction accuracy as compared to unit 3. This observation holds true even though a considerably larger number of sequences are available for unit 3 than for unit 2 across all distance ranges, as shown in Fig. \ref{fig:seq_vs_dist}. This finding further highlights that some semantic representations may be more effective in certain regions than in others.
	
	\begin{figure}[t]
		\centering
		\includegraphics[width=1\linewidth]{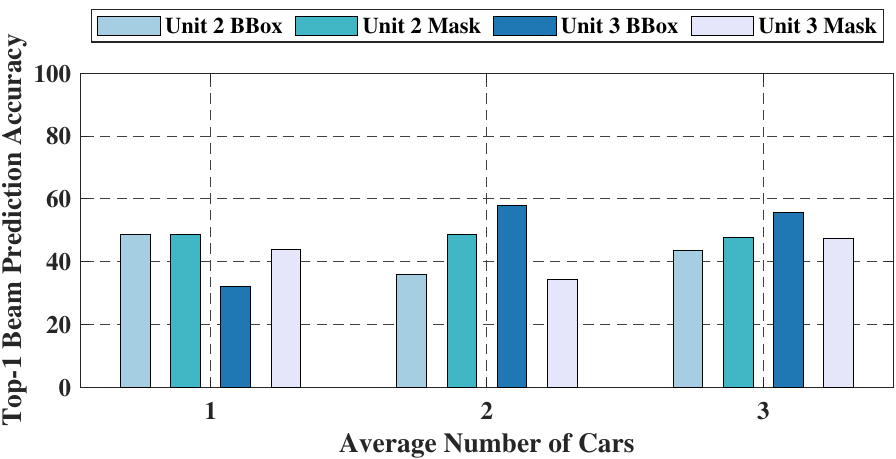}
		\caption{The figure shows how the top-1 beam prediction accuracies of the LSTM models change with the number of mobile objects present in the wireless environment for both units 2 and 3. We observe that the beam prediction accuracies remain stable and, in some cases, even improve as the average number of objects in the wireless environment increases. This highlights the robustness of the beam prediction approach in scenarios involving multiple candidates.}
		\label{fig:acc_vs_cars}
	\end{figure}
	\begin{figure}[t]
		\centering
		\includegraphics[width=1\linewidth]{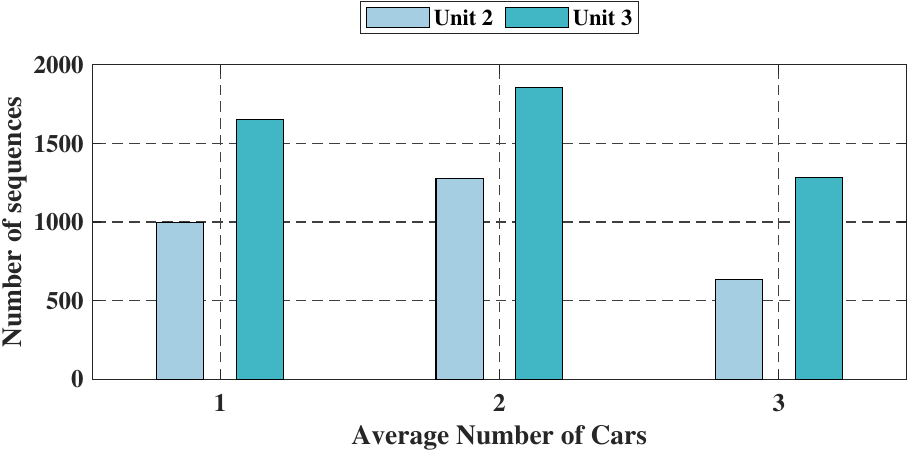}
		\caption{The figure shows the relationship between the number of image sequences and the average number of objects of interest present within those sequences. We observe a substantial difference in the number of available sequences between unit 2 and unit 3, with unit 3 having a considerably larger number of sequences.}
		\label{fig:seq_vs_cars}
	\end{figure}
	\textbf{How does the average number of objects of interest present in the wireless environment affect beam prediction accuracy?}
	
	Fig.~\ref{fig:acc_vs_cars} shows how the top-1 beam prediction accuracies from the LSTM models vary with the average number of objects of interest present in the wireless environment for both units 2 and 3. The average is determined by considering the total number of relevant objects across the five image samples in the sequence. We note that the beam prediction accuracies remain stable and even increase in some instances as the average number of objects in the wireless environment increases. This underscores the efficacy of the proposed transmitter identification and tracking solution and demonstrates the overall effectiveness of the proposed beam prediction solution in a multi-candidate scenario. Fig.~\ref{fig:seq_vs_cars} depicts the relationship between the number of image sequences and the average number of relevant objects present within those sequences. We observe a substantial difference in the number of available sequences between unit 2 and unit 3, with unit 3 having a significantly larger number of sequences across the average number of objects categories. However, we do not see a proportional increase in beam prediction accuracies for unit 3 compared to unit 2 across the average number of objects categories. Notably, in the category of only one object being present in the sequence, unit 3 has considerably larger number of sequences compared to unit 2. However, as depicted in Fig.~\ref{fig:acc_vs_cars}, both the bounding box-based and mask-based LSTM models of unit 2 exhibit superior beam prediction accuracy compared to the LSTM models of unit 3 in the category of only one object being present in the sequence.

	\section{Conclusion}\label{sec:conc}
	This paper presents a distributed sensing-aided beamforming approach. The proposed solution involves deploying multiple distributed nodes, which extract masks and bounding boxes of potential users from raw RGB images. We effectively reduce the storage and transmission requirements by transmitting these semantics to the basestation instead of raw RGB images. We also propose a transmitter identification and tracking solution at the basestation, enabling the proposed solution to operate in a multi-candidate setting. Experimental results on the DeepSense 6G dataset demonstrate the effectiveness of the proposed solution in identifying and tracking the transmitter over multiple frames. The results further show that the proposed solution can predict the optimal beam effectively and demonstrates robustness against both increasing distances from the distributed nodes and a higher number of objects of interest present in the wireless environment. These findings highlight the potential of utilizing environment semantics to facilitate distributed sensing-aided communication.
	
	\balance
	%==========
	\bibliographystyle{IEEEtran}
	%\bibliography{Ref_final}
	

\end{document}